\documentclass[a4paper,fleqn,usenatbib,useAMS]{mnras}
\makeatletter
\let\Bbbk\@undefined 
\makeatother

\usepackage[dvipsnames]{xcolor}
\usepackage{tikz,hyperref}
\usepackage{soul}

\definecolor{lime}{HTML}{A6CE39}
\DeclareRobustCommand{\orcidicon}{
	\begin{tikzpicture}
	\draw[lime, fill=lime] (0,0) 
	circle [radius=0.13] 
	node[white] {{\fontfamily{qag}\selectfont \tiny ID}};
	\draw[white, fill=white] (-0.0625,0.095) 
	circle [radius=0.007];
	\end{tikzpicture}
	\hspace{-2mm}
}

\foreach \x in {A, ..., Z}{\expandafter\xdef\csname orcid\x\endcsname{\noexpand\href{https://orcid.org/\csname orcidauthor\x\endcsname}
			{\noexpand\orcidicon}}
}

\usepackage{booktabs}

\usepackage[T1]{fontenc}
\usepackage{ae,aecompl}

\usepackage{graphicx}	
\usepackage{amsmath} 
\usepackage{amssymb} 
\usepackage{newtxtext} 
\usepackage{booktabs}
\usepackage{times}
\usepackage{float}
\usepackage{xcolor}
\input{epsf}

\title[Nova remnant morphological catalogue]
{A morphological catalogue of Nova Remnants} 
\author[E.\ Santamar\'{i}a et al.]{E.~Santamar\'{i}a\thanks{E-mail:\,e.santamaria@irya.unam.mx}$^{1}\orcidA$, M.~A.\ Guerrero$^{2}\orcidB$, G.~Ramos-Larios$^{3}\orcidC$, J.~A.~Toal\'a$^{1}\orcidD$ and L.~Sabin$^{4}\orcidE$
\\
$^{1}$Instituto de Radioastronom\'{i}a y Astrof\'{i}sica, Universidad Nacional Aut\'{o}noma de M\'{e}xico, 58090 Morelia, Michoac\'{a}n, Mexico\\
$^{2}$Instituto de Astrof\'{\i}sica de Andaluc\'{\i}a, IAA-CSIC, Glorieta de la Astronom\'{\i}a s/n, 18008 Granada, Spain\\
$^{3}$Instituto de Astronom\'{i}a y Meteorolog\'{i}a, CUCEI, Universidad de Guadalajara, Av. Vallarta 2602, Col. Arcos Vallarta, 44130 Guadalajara, Mexico \\
$^{4}$Universidad Nacional Aut\'{o}noma de M\'{e}xico, Instituto de Astronom\'{i}a, A.P. 877, 22800 Ensenada, B.C., Mexico
}

\date{\today}

\pubyear{2025}

\begin{document}
\label{firstpage}
\pagerange{\pageref{firstpage}--\pageref{lastpage}}
\maketitle

\begin{abstract}
This work presents the first optical imaging catalogue of resolved Galactic nova remnants. 
It compiles images from proprietary observations carried out at the Nordic Optical Telescope and public archives using different telescopes and instruments. 
The catalogue includes images spanning from 1950 to 2024 of 66 novae out of the more than 550 novae detected to date.
Diffuse emission from a nova shell is detected in 45 sources, with another 16 sources been stellar and five more been barely resolved. 
We used the catalogue to introduce a main morphological classification of the nova remnants (R - round, E - elliptical, B - bipolar, A - asymmetric, I - irregular, S - stellar, and G - barely resolved) with secondary features (s - smooth, c - clumpy, m - mixed, f - filaments, t - tails, m - multiple shells, e - equatorial brightness enhancement).  
Resolved nova remnants are mostly round (27\%) or elliptical (56\%), with very few bipolar or with equatorial brightness enhancement (13\%). 
Most nova remnants are younger than 150 yr, with about 80\% of nova remnant detection rate among those younger than 60 yr. 
The physical size of nova remnants increases with age at a rate of 0.0725 pc per century up to a median value of 0.03 pc.  
This relationship seems to apply even to the few known ancient nova shells with ages up to a few millennia.  
The aspect ratio of nova remnants can be described by a Gaussian distribution with a standard deviation $\sigma$ of 0.18 from round morphologies (major to minor axes ratio of unity). 
\end{abstract}




\section{Introduction}\label{introduction}

Accreting white dwarfs (WDs) exhibit a variety of phenomena depending on the nature of their mass-losing companions \citep{Mukai2017}. They can be used as laboratories for studying mass accretion, stellar evolution and the enrichment of the interstellar medium (ISM) at the end points of the lives of Solar-like stars, as their final destiny might be type Ia supernova explosions if they reach the Chandrasekhar limit.

Among the systems that involve an accreting WD are cataclysmic variables (CVs). In such cases, a WD accretes material from a late star on or near the main sequence that fills its Roche lobe \citep{Bode2008}. 
CVs produce classical novae (CNe) events when the accreted material at the surface of the WD reaches critical conditions to produce a thermonuclear runaway \cite[$T_\mathrm{crit} \approx10^{7}$~K, $P_\mathrm{crit} \approx10^{20}$~dyn~cm$^{-2}$;][]{Truran1986,Gehrz1998,Starrfield2016}. 
These violent events inject about 10$^{-6}$--10$^{-4}$~M$_\odot$ into the ISM at high velocities \citep{Shafter2002,Bode2010,Wolf2013}. The CN events are characterised by an initial slow wind (500--2000~km~s$^{-1}$) followed by a faster one \citep[1000--4000~km~s$^{-1}$;][]{Bode2008}. 
The CN event and subsequent wind-wind interaction \citep[e.g.,][]{OBrien1994} produce what is is known as a nova remnant.

\begin{table*}
\centering
\caption{Telescopes, cameras and filters used to obtain the images of nova remnants presented in this compilation. }
\label{tab:tel}
\resizebox{\textwidth}{!}{%
\begin{tabular}{lclcccl}
\hline
\hline

\multicolumn{1}{l}{Telescope} &
\multicolumn{1}{c}{Aperture} & 
\multicolumn{1}{l}{Instrument} &
\multicolumn{1}{c}{Plate scale} & 
\multicolumn{1}{c}{Filter Name} &
\multicolumn{1}{c}{Filter Width} &
\multicolumn{1}{l}{Instrument Code} \\

\multicolumn{1}{c}{} & 
\multicolumn{1}{c}{(m)} & 
\multicolumn{1}{c}{} & 
\multicolumn{1}{c}{(arcsec~pix$^{-1}$)} & 
\multicolumn{1}{c}{} & 
\multicolumn{1}{c}{(\AA)} & 
\multicolumn{1}{c}{} \\

\hline
{\it Hubble Space Telescope} & 2.4 & Wide Field Planetary Camera 2 & 0.05 & F656N, F658N & 28, 39 & HST WFPC2 \\
{\it Hubble Space Telescope} & 2.4 & Wide Field Camera 3 & 0.04 & F502N, F657N, F680N & 65, 121, 28 & HST WFC3 \\
Nordic Optical Telescope & 2.5 & Alhambra Faint Object Spectrograph and Camera & 0.21 & OSN H01 H$\alpha$ & 10 & NOT ALFOSC n \\
Nordic Optical Telescope & 2.5 & Alhambra Faint Object Spectrograph and Camera & 0.21 & NOT \#21 H$\alpha$ & 33 & NOT ALFOSC m \\
Palomar Observatory & 1.2 & Photographic Plate &  1.70 & 103aE (Red) & 343 & PO \\
Very Large Telescope UT4 & 8.2 & Multi Unit Spectroscopic Explorer (NFM) & 0.20 & H$\alpha$ & $\dots$ & VLT MUSE \\
Anglo-Australian Telescope & 3.9 & TAURUS-2 & 0.32 & AAO \#57 H$\alpha$ & 45 & AAT TAURUS-2 \\
Kitt Peak National Observatory & 4.0 & Mosaic CCD camera & 0.26 & H$\alpha$ k1009 & 83 & KPNO CCD \\
William Herschel Telescope & 4.2 & Auxiliary-port CAMera (ACAM) & 0.25 & T6565 & 23 & WHT ACAM \\
New Technology Telescope & 3.6 & EFOSC & 0.24 & ESO \#692 H$\alpha$ & 64 & NTT EFOSC \\
\hline
\end{tabular}
}
\vspace{0.15cm}
\end{table*}

Characterising nova remnants can help us to peer into the nova event itself. The final morphology of a nova shell is impacted by the complex interaction between the binary components, the accretion disc and the explosion morphology. For years, the main discussed characteristic was the production of prolate or oblate ellipsoidal morphologies \citep{Lloyd1993,Lloyd1997,Mustel1970}. 
Numerical simulations have unveiled that nova remnants can be broadly described by prolate ellipsoidal shapes \citep{Walder2008,Orlando2017}, with clumps and filaments resulting from complex instabilities in the wind-wind interaction region.

The General Catalogue of Variable Stars (GCVS, 4th edition \citep{Kholopov+1985a,Kholopov+1985b,Kholopov+1987} is a foundational reference for variable star data. \citet{Duerbeck+1987} supplemented this with an enhanced catalogue focused on novae, listing 277 objects, including CNe and related objects (recurrent novae, X-ray novae, dwarf novae with long cycle lengths, symbiotic stars, and suspected new stars). The Catalogue and Atlas of Cataclysmic Variables \citep[1st and 2nd Editions;][]{Downes+1993,Downes+1997} has been a valuable source of information for the CV community. \citet{Downes+2001} developed a ``living'' edition of the catalogue, an evolving version that integrates orbital period data and finding charts for novae.
\cite{Ritter+2003} compiled a catalogue with detailed coordinates, magnitudes, orbital parameters, and stellar characteristics for 472 CVs, 71 low-mass X-ray binaries, and 113 related objects. It also included references to finding charts and cross-references for object aliases. 
More recently, \cite{Saito+2013} added a near-infrared catalogue of novae in the VISTA Variables in the Via Lactea (VVV) survey area, documenting $J$, $H$ and $K$ colours for 93 objects from 140 catalogued novae.

\begin{figure}
\centering
\includegraphics[width=1.0\linewidth]{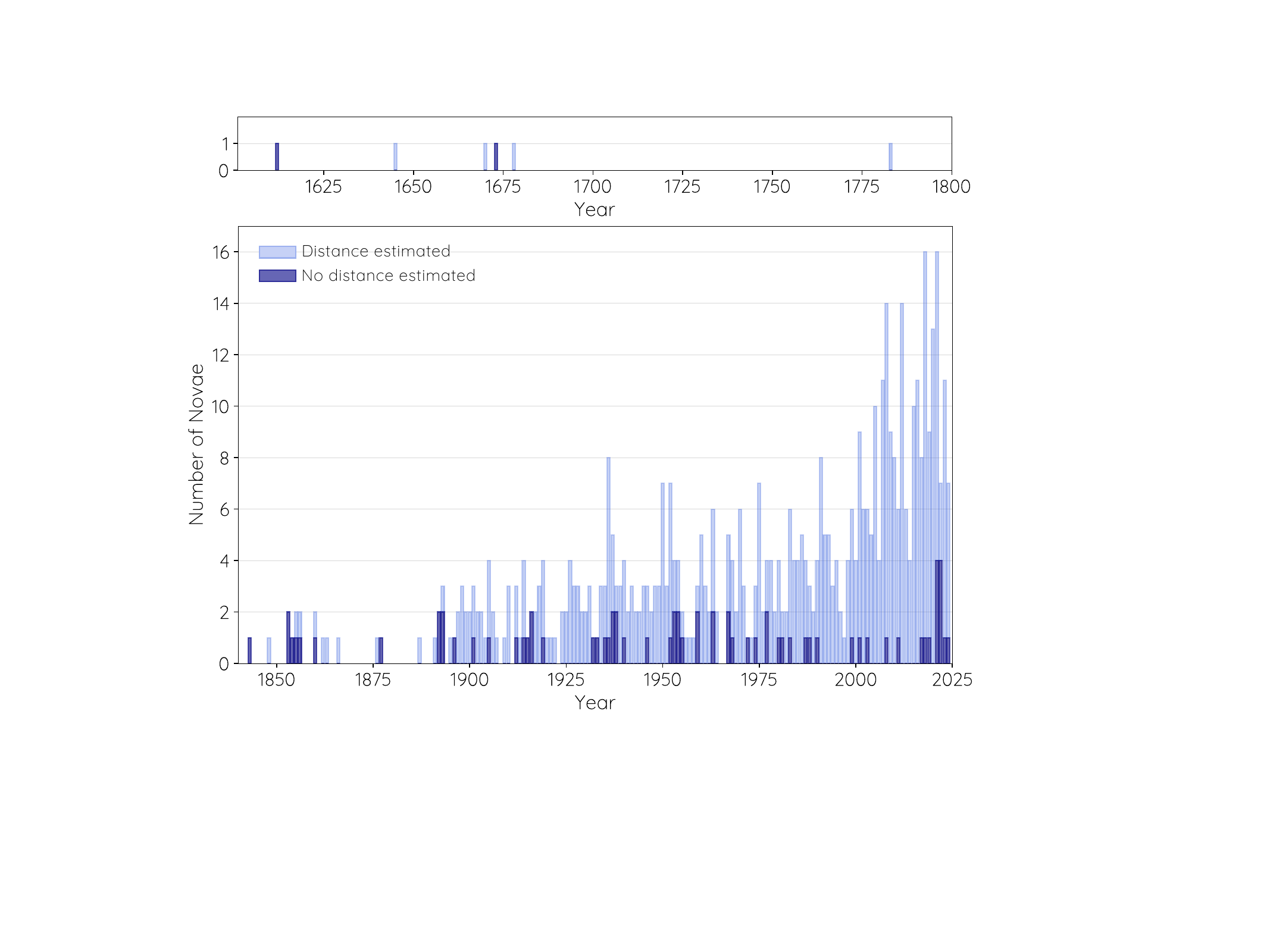}
\caption{
Yearly number of novae detected since 1612, from the early seventeenth century until 2024. 
In the histogram, dark blue bars correspond to novae with no distance estimate and pale blue bars to novae with distance estimate \citep{Bailer2021,Schaefer2022}.
}
\label{fig:year}
\end{figure}

Unlike supernova remnants, which expand into larger regions of the ISM and are easily observable across various wavelengths, nova remnants are smaller and more challenging to detect. 
This difficulty is compounded by their tendency to blend into the surrounding ISM \citep{2024arXiv}. Short exposure times hinder the detection of nova remnants, making them challenging objects to study. Nova outbursts are relatively brief, typically lasting a few weeks to months \citep{Starrfield2016}, and the remnants diminish rapidly after the explosion. 
This rapid temporal evolution, occurring on a human timescale, necessitates precise and timely observations \citep{Santa+2020}. 
Once the nova fades, detecting and characterising its remnant requires advanced instrumentation or extended monitoring efforts \citep{Bode2008}.

Despite these challenges, studying nova remnants is crucial for capturing their defining characteristics \citep{Bode2008}. 
The only large-scale effort to image nova remnant was presented by \cite{Downes+2000}, which included optical imaging of 30 recent novae.  
Resolved shells were identified around nine objects via ground-based observations, and another four using {\it Hubble Space Telescope (HST)} observations. In this paper we attempt to create the first ever imaging catalogue of nova remnants around CVs. 
The core of this catalogue is based on extensive proprietary observations obtained during the past years. 
In addition, we made an exhaustive search in a large number of telescope archives to complement our targets. 
A morphological classification is proposed that is enhanced with statistics on the characteristics of nova remnants.

This paper is organized as follows. In Section~2 we present our sample and provide details on the catalogues explored for observations. A classification scheme is presented in Section~3, along with a description of the general properties of the objects in the sample. We present our conclusions in Section~4.

\begin{table*}
\centering
\caption{Novae with resolved shells. The objects are arranged by constellation name.}
\label{tab:nov}
\resizebox{\textwidth}{!}{%
\begin{tabular}{lllrllrlccccc}
\hline
\hline
\multicolumn{1}{l}{Object} &
\multicolumn{1}{c}{$l, b$} & 
\multicolumn{1}{l}{Speed Class} &
\multicolumn{1}{c}{$t_{3}$} &
\multicolumn{1}{c}{Outburst Date} & 
\multicolumn{1}{c}{$d$}  & 
\multicolumn{1}{c}{$z$}  & 
\multicolumn{1}{c}{Instrument} & 
\multicolumn{1}{c}{REACH} & 
\multicolumn{1}{c}{Image Epoch} & 
\multicolumn{1}{c}{FWHM$_\mathrm{inst}$} &
\multicolumn{1}{c}{Morphology} & 
\multicolumn{1}{c}{Angular Radius}  \\

\multicolumn{1}{c}{} &
\multicolumn{1}{c}{($^\circ$)} & 
\multicolumn{1}{c}{} & 
\multicolumn{1}{c}{(day)} & 
\multicolumn{1}{c}{} & 
\multicolumn{1}{c}{(pc)}  & 
\multicolumn{1}{c}{(pc)}  & 
\multicolumn{1}{c}{} & 
\multicolumn{1}{c}{(cm$^{2}$ s)} & 
\multicolumn{1}{c}{} & 
\multicolumn{1}{c}{(")} & 
\multicolumn{1}{c}{}  &
\multicolumn{1}{c}{(")} 
\\ 
\hline
OS\,And$^{\star}$    & 106.05$-$12.11 & fast            &  23~~ &  ~~1986 Dec & 4152$^{+1177}_{-603}$ & -870~ & HST WFPC2 & $2.05\times10^{7}$ & 1998 Mar& 0.1 & R & 0.3 \\
DO\,Aql    &  31.70$-$11.80 & very slow            & 900~~ &  ~~1925 Sep & 4357$^{+1207}_{-646}$ & -890~ & NOT ALFOSC m &$1.24\times10^{8}$ & 2020 Aug& 0.9 & I & $2.5\times3.4$ \\
V603\,Aql  & 33.16$+$00.82  & very fast       &  12~~ &  ~~1918 Jun & 315$^{+3}_{-3}$   & 5~  & PO and 103aE & $\dots$ & 1950 Jul& $\dots$ & E & $25\times28$ \\
V1229\,Aql$^{\star}$ &  40.53$-$05.43 &  fast  &  32~~ &  ~~1970 Apr & 3317$^{+1415}_{-583}$ & -310~  & VLT MUSE&$7.60\times10^{8}$ & 2022 Sep& 0.8 & Es & $2.1\times2.7$ \\
V1425\,Aql & 33.01$-$03.89  & moderately fast  &  79~~ &  ~~1995 Feb & 5711$^{+2164}_{-1444}$ & -385~ & HST WFPC2& $3.62\times10^{5}$ & 1997 Nov& 0.1 & R & $0.24$ \\
V1494\,Aql &  40.97$-$04.74 & very fast       &  16~~ &  ~~1999 Dec & 858$^{+153}_{-74}$ & -70~ & NOT ALFOSC m&$1.07\times10^{8}$ & 2022 Aug& 0.7 & Rs & 5.4 \\
T\,Aur     & 177.14$-$01.70 & moderately fast &  84~~ &  ~~1891 Dec & $802^{+44}_{-26}$   &  -25~  & NOT ALFOSC n&$8.02\times10^{7}$ & 2016 Nov& 0.8 & Ect & $10.0\times13.6$\\
V842\,Cen  & 316.57$+$2.45  & fast            &  48~~ &  ~~1986 Nov & 1374$^{+89}_{-64}$  &  58~  & HST WFPC2& $1.96\times10^{8}$ & 1999 Mar& 0.1 & Ecft & $1.15\times1.23$\\
AT\,Cnc    & 198.54$+$31.74 & $\dots$             &  $\dots$~ &  ~~1645-1686 & 455$^{+7}_{-5}$  & 240~ & KPNO CCD&$6.40\times10^{8}$ & 2010 Feb& 1.3 & Ic & $47\times110$ \\
CP\,Cru    & 298.11$+$00.73 & very fast       &  10~~ &  ~~1996 Aug & 3570$^{+1329}_{-921}$ & 45~ & HST WFPC2& $2.05\times10^{7}$ & 1997 Nov&0.1 & I/Ee & $0.20\times0.25$  \\
V476\,Cyg  &  87.36$+$12.41 & very fast       & 16~~ &  ~~1920 Aug & 1228$^{+209}_{-154}$ & 260~ & NOT ALFOSC n&$4.63\times10^{7}$ & 2018 Jun& 0.7 & E & $13.2\times14.6$ \\
V1500\,Cyg &  89.82$-$00.07 & very fast       &   4~~ &  ~~1975 Aug & 4345$^{+718}_{-710}$ & -5~ & WHT ACAM&$1.66\times10^{8}$ & 1993 Aug& 1.0 & Ec & $5.09\times5.56$ \\
V1819\,Cyg &  71.37$+$03.97 & slow            & 181~~ &  ~~1986 Aug & 18303$^{+3406}_{-2040}$ & 1265~ & HST WFPC2& $1.36\times10^{7}$ & 1999 Mar&0.1 & E & $0.18\times0.20$  \\
V1974\,Cyg &  89.13$+$07.81 & fast            &  43~~ &  ~~1992 Feb & 1631$^{+236}_{-122}$ & 220~ & NOT ALFOSC m&$1.07\times10^{8}$ & 2020 Aug& 0.7 & Ect & $4.4\times5.0$ \\
HR\,Del    &  63.43$-$13.97 & slow            & 231~~ &  ~~1967 Jul & 872$^{+14}_{-14}$ & -210~ & HST WFPC2& $1.80\times10^{8}$ & 1997 May&0.1 & Ecft  & $3.4\times4.3$  \\
DQ\,Her    &  73.15$+$26.44 & moderately fast & 100~~ &  ~~1934 Dec & 491$^{+5}_{-3}$  &  220~  & NOT ALFOSC n&$8.02\times10^{7}$ & 2017 May&0.9 & Ecft  & $12.8\times18.3$  \\
V533\,Her  &  69.18$+$24.27 & fast            &  43~~ &  ~~1963 Feb & 1250$^{+44}_{-35}$ &  510~ & NOT ALFOSC n&$6.18\times10^{7}$ & 2019 Jun& 1.1 & Rr & $9.0$  \\
J210204    &  88.01$+$00.45 & $\dots$             &  $\dots$~ & ~~1850-1890 & 600$^{+130}_{-130}$ & 5~ & NOT ALFOSC n&$4.63\times10^{7}$ & 2017 Jun& 0.6 & Ic & $25\times34$ \\
DK\,Lac    & 105.23$-$05.35 & slow            & 202~~ &  ~~1950 Jan & 3159$^{+365}_{-416}$ & -295~ & WHT ACAM&$1.25\times10^{8}$  & 1993 Sep& 0.9 & Es & $1.9$ \\
BT\,Mon    & 213.86$-$02.62 & slow            & 182~~ &  ~~1939 Dec & 1498$^{+89}_{-61}$ &  -70~  & NOT ALFOSC m&$1.51\times10^{8}$ & 2021 Feb&1.0 & E & $7.5\times9.0$ \\
V959\,Mon  & 206.34$+$00.07 & $\dots$             & $\dots$~   &  ~~2012 Aug & 3388$^{+791}_{-680}$ & 5~ & HST WFC3&$9.48\times10^{7}$ & 2015 Nov& 0.1 & Bec & $0.50\times0.75$ \\
V2264\,Oph$^{\star}$ & 358.77$+$05.86 & fast            &  30~~ &  ~~1991 Apr & 8155$^{+738}_{-738}$ & 830~ & HST WFPC2& $5.43\times10^{6}$ & 1999 Mar& 0.1 & Rc & $0.80$ \\
GK\,Per    & 150.95$-$10.10 & very fast       &  13~~ &  ~~1901 Feb & 431$^{+9}_{-7}$    & -75~ & NOT ALFOSC m&$8.02\times10^{7}$ & 2020 Aug& 0.8 & Ect & $51\times58$ \\
RR\,Pic    & 272.35$-$25.67 & moderately fast & 122~~ &  ~~1925 May & 491$^{+5}_{-6}$   & -210~ & AAT TAURUS-2&$1.08\times10^{8}$ & 1995 Feb& 0.9 & Ecte & $12.4\times14.0$\\
CP\,Pup    & 252.92$-$00.83 & very fast       &   8~~ &  ~~1942 Nov & 758$^{+10}_{-10}$  & -10~ & AAT TAURUS-2&$2.15\times10^{8}$ & 1995 Feb& 1.2 & Rc & $8.9$ \\
DY\,Pup$^{\star}$   & 245.82$+$04.36 & slow            & 160~~ &  ~~1902 Nov & 3585$^{+2466}_{-794}$  & 270~ & AAT TAURUS-2& $2.15\times10^{8}$& 1995 Feb& 0.9 & E & $2.8\times3.6$\\
V351\,Pup  & 252.72$-$00.73 & fast            &  26~~ &  ~~1991 Dec & 12048$^{+1834}_{-2343}$ &  -155~ & HST WFPC2& $1.02\times10^{7}$ & 1998 Feb&0.1 & Rmx & $0.51\times0.63$  \\
V445\,Pup$^{\star}$  &  241.12$-$02.19   & slow & 240~~ &  ~~2000 Dec  & 6272$^{+2754}_{-1246}$  & -240~ & HST WFC3&$2.07\times10^{8}$ & 2015 Jan& 0.1 & B & $0.86\times1.38$     \\
T\,Pyx     & 257.20$+$09.70 & moderately fast  &  62~~ & ~~2011 Apr         & 2589$^{+262}_{-148}$ &  435~ & HST WFPC2&$1.88\times10^{8}$ & 1994 Feb& 0.1 & Rc & $4.7$ \\
V960\,Sco  & 358.71$-$03.50 & $\dots$         &  $\dots$~ &  ~~1985 Sep & 6961$^{+2336}_{-1763}$ &  -425~ & HST WFPC2&$2.05\times10^{7}$ & 1998 Oct& 0.1 & E & $0.17\times0.22$ \\
V992\,Sco  &  343.82$-$1.60 &  moderately fast  &  120~~  &  ~~1992 May  & 1783$^{+400}_{-258}$  & -50~ & HST WFPC2&$1.36\times10^{7}$ & 1998 Aug& 0.1 & R & $0.3$  \\
CT\,Ser    &  24.48$+$47.56 &  moderately fast  & 100~~  &  ~~1948 Apr & 4029$^{+996}_{-699}$ &  2970~ & NOT ALFOSC n&$6.18\times10^{7}$ & 2018 Jun& 0.8 & Rc & $4.5$ \\
FH\,Ser    &  32.90$+$05.78 & moderately fast &  62~~ &  ~~1960 Feb & 1084$^{+74}_{-92}$ & 110~ & HST WFPC2&$3.07\times10^{8}$ & 1997 May& 0.1 & Ecef & $6.1\times7.1$ \\
V4077\,Sgr &  07.42$-$08.31 & moderately fast & 100~~ &  ~~1982 Oct & 9658$^{+2055}_{-1391}$ & -1395~ & HST WFPC2& $1.02\times10^{7}$& 1999 Mar&0.1 & Ee & $0.24\times0.35$  \\
V4121\,Sgr$^{\star}$ &  02.54$-$04.16 & very fast      & 15~~    &  ~~1983 Feb & 8135$^{+739}_{-748}$ & -590~ & HST WFPC2&$1.79\times10^{7}$ & 1999 Mar& 0.1 & I & $0.32$ \\
V5668\,Sgr$^{\star}$ & 05.38$-$09.86 &  moderately fast & 78~~   &  ~~2015 Mar & 1290$^{+245}_{-585}$ & -220~ & HST WFC3&$1.81\times10^{6}$ & 2017 Jul& 0.1 & Em & $0.26\times0.32$ \\
XX\,Tau    & 187.10$-$11.65 & fast            &  42~~ &  ~~1927 Nov & 7498$^{+2020}_{-1427}$ & -1515~ & AAT TAURUS-2&$7.17\times10^{6}$ & 1995 Feb&1.2 & I & $1.9$  \\
RW\,UMi    & 109.63$+$33.15 & moderately fast & 140~~ &  ~~1956 Sep & 2260$^{+796}_{-550}$ & 1235~ & WHT ACAM&$8.31\times10^{7}$ & 1993 Sep& 1.1 & R & $1.3$ \\
V382\,Vel  & 284.16$+$05.77 & very fast       &  13~~ &  ~~1999 May & 1497$^{+92}_{-101}$ & 150~ & NTT EFOSC& $4.49\times10^{7}$& 2017 Apr&0.6 & Ec & $3.0\times3.6$  \\
LV\,Vul$^{\star}$    &  63.30$+$00.84 & fast            &  38~~ &  ~~1968 May & 2887$^{+1347}_{-467}$ & 40~ & WHT ACAM&$1.25\times10^{8}$ & 1993 Sep&1.0 & Rs & $2.5$ \\
NQ\,Vul    &  55.35$+$01.28 & moderately fast &  50~~ &  ~~1976 Oct & 1180$^{+116}_{-86}$ & 25~ & NOT ALFOSC m&$1.07\times10^{8}$ & 2020 Aug& 0.9 & Es & $4.9\times5.3$ \\
PW\,Vul    &  61.09$+$05.19 & moderately fast & 116~~ &  ~~1984 Jul & 2185$^{+528}_{-305}$ & 200~ & NOT ALFOSC m&$1.07\times10^{8}$ & 2020 Aug& 0.9 & Es & $2.0\times2.5$ \\
QU\,Vul    &  68.51$-$06.02 & fast            &  36~~ &  ~~1984 Dec &  1430$^{+230}_{-230}$ & -150~ & HST WFPC2&$2.05\times10^{7}$ & 1998 Jan& 0.1 & Efmx & $0.74\times0.97$ \\
QV\,Vul    &  53.85$+$06.97 & fast            &  47~~ &  ~~1987 Nov & 5417$^{+1199}_{-955}$ & 660~ & NOT ALFOSC m&$1.07\times10^{8}$ & 2020 Jul& 1.0 & Es & $1.42\times1.66$ \\
V458\,Vul  &  58.63$-$03.61 & very fast  &  20~~ &  ~~2007 Aug & 10487$^{+1344}_{-999}$ & -660~ & NOT ALFOSC n&$5.35\times10^{7}$ & 2024 Jul& 0.6 & Ext & $11.3\times15.8$  \\
\hline
\multicolumn{9}{@{}p{\linewidth}@{}}{\footnotesize Note: $d$ corresponds to distances estimated by \citet{Bailer2021}. $^{\star}$References to distances were obtained from \citet{Schaefer2022}.}
\end{tabular}
}
\vspace{0.15cm}
\end{table*}


\begin{table*}
\centering
\caption{Novae without nebular emission (undetected/stellar) arranged by constellations.}
\label{tab:nov_st}
\resizebox{\textwidth}{!}{%
\begin{tabular}{lllrlllccccc}
\hline
\hline
\multicolumn{1}{l}{Object} &
\multicolumn{1}{c}{$l, b$} & 
\multicolumn{1}{l}{Speed Class} &
\multicolumn{1}{c}{$t_{3}$} &
\multicolumn{1}{c}{Outburst Date} & 
\multicolumn{1}{c}{$d$}  & 
\multicolumn{1}{c}{Instrument} & 
\multicolumn{1}{c}{REACH} & 
\multicolumn{1}{c}{Image Epoch} & 
\multicolumn{1}{c}{FWHM$_\mathrm{inst}$} &
\multicolumn{1}{c}{Morphology} & 
\multicolumn{1}{c}{Angular Radius}  \\
\multicolumn{1}{c}{} &
\multicolumn{1}{c}{($^\circ$)} & 
\multicolumn{1}{c}{} & 
\multicolumn{1}{c}{(day)} & 
\multicolumn{1}{c}{} & 
\multicolumn{1}{c}{(pc)}  & 
\multicolumn{1}{c}{} & 
\multicolumn{1}{c}{(cm$^{2}$ s)} & 
\multicolumn{1}{c}{} & 
\multicolumn{1}{c}{(")} & 
\multicolumn{1}{c}{} & 
\multicolumn{1}{c}{(")} 
\\  
\hline
AP\,Cru           & 300.76$-$01.65 & $\dots$          & $\dots$~ & ~~1935 Mar &  3672$^{+794}_{-578}$   & NTT EFOSC &   $3.05\times10^{7}$& 2017 Apr& 0.74 & G  &    0.10  \\ 
HY\,Lup           & 318.53$+$08.62 & fast             &     26~~  & ~~1993 Sep &  3814$^{+2444}_{-2122}$ & NTT EFOSC &   $6.11\times10^{7}$& 2017 Jul& 1.3 & G  &    0.31   \\ 
GQ\,Mus  &  297.21$-$04.99 & fast   & 45~~  &  ~~1983 Jan &  16968$^{+4131}_{-2582}$  &  NTT EFOSC&$1.83\times10^{7}$ & 2017 Apr& 0.75 & G & 0.40  \\ 
IL\,Nor$^{\star}$ & 326.83$+$04.81 & moderately fast  &     108~~ & ~~1893 Jul & 3318$^{+2562}_{-914}$   & AAT TAURUS-2& $2.15\times10^{8}$& 1995 Feb& 0.6 & G & 0.50   \\
HS\,Pup           & 247.75$-$02.11 & fast             &      65~~ & ~~1963 Dec &  4409$^{+1925}_{-880}$  & AAT TAURUS-2& $2.15\times10^{8}$& 1995 Feb& 0.82 & G  &    0.28  \\ 
\hline
V500\,Aql  & 047.60$-$09.46  &  fast      & 42~~  & ~~1943 Apr  &  5159$^{+1709}_{-3133}$   & NTT EFOSC&$6.11\times10^{7}$ & 2017 Jul& 1.2  & S & $\dots$   \\
RS\,Car  & 291.09$-$01.46  &  moderately fast   &  80~~  & ~~1895 Apr  &  12310$^{+2604}_{-5426}$   & AAT TAURUS-2&$2.15\times10^{8}$ &  1995 Feb& 0.9 & S & $\dots$   \\
V365\,Car & 289.16$+$01.49 &  very slow       & 530~~ & ~~1948 Jul  &  4007$^{+956}_{-730}$   & AAT TAURUS-2&$1.08\times10^{8}$ & 1995 Feb& 1.7 & S & $\dots$   \\
RR\,Cha   & 304.16$-$19.54 & moderately fast  &  60~~ &  ~~1953 Apr & 6115$^{+1652}_{-934}$   & AAT TAURUS-2& $2.15\times10^{8}$& 1995 Feb& 1.2 & S & $\dots$  \\
AR\,Cir & 317.03$-$00.37  & very slow   &  330~~  & ~~1906 Feb  &  1496$^{+165}_{-166}$   & NTT EFOSC&$6.11\times10^{6}$ & 2017 Jul& 1.3  & S & $\dots$  \\
MT\,Cen & 294.73$+$01.23  &  very fast  &  10~~ & ~~1931 May  &  2493$^{+882}_{-578}$   & AAT TAURUS-2&$7.17\times10^{6}$ & 1995 Feb& 0.7  & S & $\dots$   \\
V359\,Cen & 292.41$+$20.01  & $\dots$  & $\dots$~  & ~~1930 Apr  &  345$^{+20}_{-20}$   & AAT TAURUS-2& $1.08\times10^{8}$& 1995 Feb& 1.7  & S & $\dots$   \\
V450\,Cyg  &  79.12$-$06.45 & moderately fast & 108~~ &  ~~1942 Sep & 4224$^{+917}_{-610}$    & WHT ACAM&$1.66\times10^{8}$ & 1993 Sep& 1.0 & S & $\dots$  \\
V446\,Her & 045.40$+$04.70 & fast       &  42~~ &  ~~1960 Mar & 1281$^{+135}_{-107}$    & NOT ALFOSC m&$1.07\times10^{8}$ & 2020 Aug& 0.8  & S & $\dots$    \\
GI\,Mon  & 222.93$+$04.74  &  fast     &  23~~ & ~~1918 Feb  & 3724$^{+450}_{-413}$  &  AAT TAURUS-2&$1.08\times10^{8}$ & 1995 Feb& 1.0  & S & $\dots$   \\
RS\,Oph  & 019.79$+$10.37 &  very fast  & 14~~  & ~~2006 Feb  &  2441$^{+209}_{-222}$ &  NOT ALFOSC n&$5.57\times10^{6}$ & 2020 Aug& 1.1 & S & $\dots$  \\
V841\,Oph & 007.62$+$17.77  & moderately fast  & 130~~  & ~~1848 Apr  &  813$^{+13}_{-13}$  & AAT TAURUS-2&$7.17\times10^{6}$ & 1995 Feb& 0.7 & S & $\dots$   \\
HZ\,Pup & 246.17$+$01.38  & fast     &  70~~ &  ~~1963 Jan &  4263$^{+789}_{-452}$  &  AAT TAURUS-2&$7.17\times10^{6}$ & 1995 Feb& 1.5 & S & $\dots$  \\
V3888\,Sgr  & 009.08$+$04.65  & very fast   & 14~~  & ~~1974 Oct  &  4463$^{+1120}_{-896}$  & NOT ALFOSC m&$1.07\times10^{8}$ & 2020 Aug& 1.3 & S & $\dots$   \\
CN\,Vel  & 287.43$+$05.17  & very slow   & 800~~ & ~~1905 Dec  &  3682$^{+4566}_{-475}$  & NTT EFOSC& $1.83\times10^{7}$& 2017 Apr& 0.8 & S & $\dots$  \\
CQ\,Vel$^{\star}$ & 272.33$-$04.89  &  fast   &  50~~ & ~~1940 Apr  &  3061$^{+2425}_{-689}$  &  AAT TAURUS-2&$1.08\times10^{8}$ & 1995 Feb& 0.9 &  S  & $\dots$  \\
\hline
\multicolumn{9}{@{}p{\dimexpr\linewidth}@{}}{\footnotesize Note: $d$ corresponds to distances estimated by \citet{Bailer2021}. $^{\star}$References to distances were obtained from \citet{Schaefer2022}.} 
\end{tabular}
}
\vspace{0.15cm}
\end{table*}

\begin{figure*}
\centering
\includegraphics[width=0.7\linewidth]{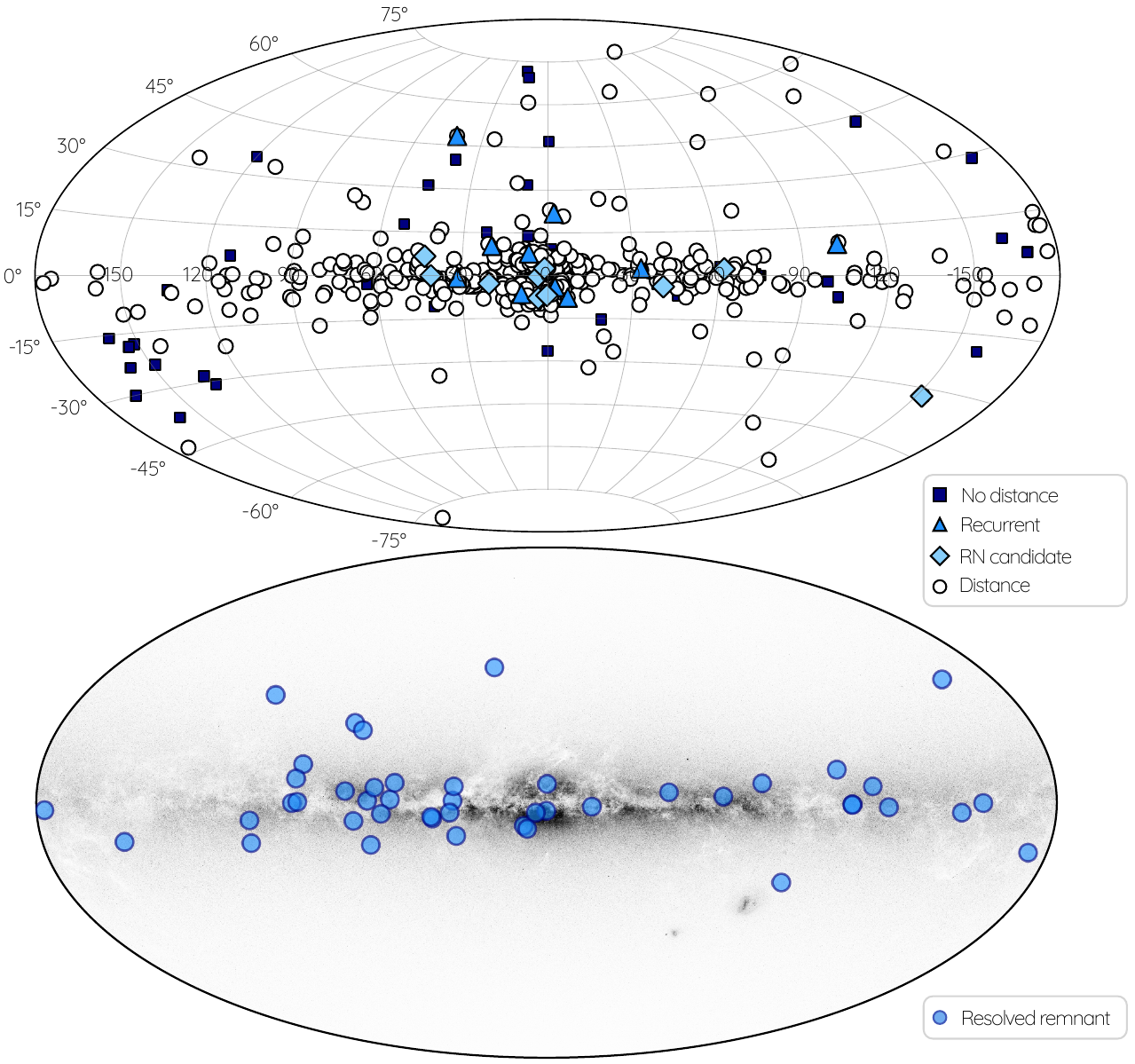}
\caption{
(top) Galactic distribution of the novae detected to date according to their Galactic coordinates. 
As in Fig.~\ref{fig:year}, novae with known distance estimates are represented by white circles, whereas blue squares are used for those with unknown distances. The triangles and diamonds indicate the position of bona-fide and candidate recurrent novae \citep{Pagnotta2014}, respectively. (bottom) The blue circles represent the position of the 45 novae with resolved nebular remnants.
}
\label{fig:aitoff}
\end{figure*}

\begin{figure*}
\centering
\includegraphics[width=1.0\linewidth]{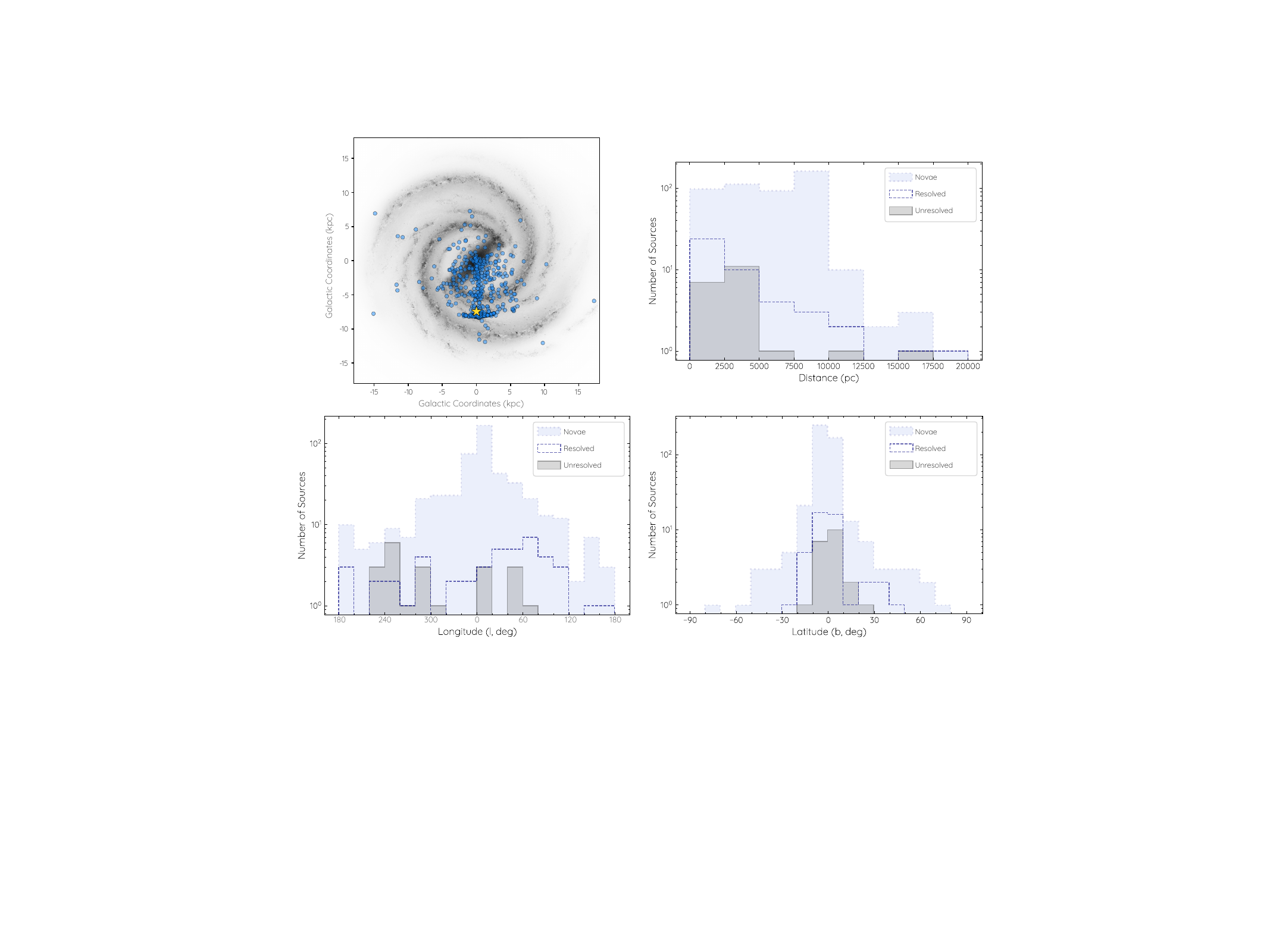}
\caption{
(top-left panel) Spatial distribution of Novae in the Galaxy. The yellow star indicates the position of the Sun. (top-right panel) The histogram shows that the highest concentration of novae is observed at a distance of 10 kpc. Histograms on $l$ and $b$ to confirm the higher frequency of occurrence on the Galactic Plane and at the direction of the Galactic Center (bottom panels). In each histogram, the light blue color represents the number of novae detected up to 2024, while the dotted blue and gray lines indicate novae with resolved and unresolved remnants, respectively.}
\label{fig:distrib}
\end{figure*}

\section{Nova remnants with available images}

Improvements in sky time coverage and survey capabilities have resulted in an increasing detection rate of nova events, which is close to exponential as plainly illustrated in Fig.~\ref{fig:year}. 
The list of known novae compiled in Koji’s List of Recent Galactic Novae\footnote{\url{https://asd.gsfc.nasa.gov/Koji.Mukai/novae/novae.html}} has been used to search for images of nova remnants through observations carried out at the Nordic Optical Telescope (NOT) at Observatorio del Roque de los Muchachos (ORM) in the island of La Palma (Spain) and among all available telescope archives. 
This search resulted in images for 66 novae using the telescopes, cameras, and filters listed in Table~\ref{tab:tel}.  
In most cases, the filters can be considered to have a narrow spectral transmission band-pass, but this is not the case for the oldest observations from Palomar Observatory, which were obtained using photographic plates with broad-band sensitive emulsions.  
The transmission band-pass of the narrow-band filters used here includes either the H$\alpha$ emission line alone or more normally includes also the [N~{\sc ii}] $\lambda\lambda$6548,6584 \AA\ emission lines, given the fast expansion velocity of nova remnants \citep[e.g.,][]{Santamaria2024}. 
Notable exceptions are images resulting from integral field spectroscopic observations, such as those obtained with the Multi Unit Spectroscopic Explorer (MUSE) at the Very Large Telescope (VLT).  
These images have a wavelength range that fits the H$\alpha$ emission line.  
An instrument code is listed in the last column of Table~\ref{tab:tel} for future reference throughout the paper.

The inspection of the available images of nova remnants either shows a resolved nova remnant or fails to detect any diffuse emission. 
The latter indicates that the nova remnant is either unresolved or has a surface brightness below the detection limit of the available image. Consequently, these objects will be referred to as stellar/undetected throughout this work.

The lists of 45 resolved and 21 stellar/undetected nova remnants are compiled in Tables~\ref{tab:nov} and \ref{tab:nov_st}, respectively.  
In both tables, columns \#1 to \#10 provide the nova name (ordered by constellation name), its Galactic coordinates, the nova type, the time in days since the outburst that it took to decline 3 mag ($t_3$), the outburst date, the distance, height above the Galactic plane ($z$), the code of the instrument used to obtain the image as listed in the last column of Table~\ref{tab:tel}, the image reach and epoch, and the full-width at half-maximum (FWHM) of stars in the field of view (i.e., the spatial resolution of the images or instrumental FWHM$_\mathrm{inst}$), respectively. The reach of an image is defined as the product of the telescope aperture and the integration time, thus providing an estimate of the image sensitivity under the assumption that the filter transmission and CCD quantum efficiency are similar for all images.

The availability of {\it Gaia} parallax measurements of the nova progenitors has allowed the determination of their distances, which is an essential parameter to derive key properties of the nova shells such as their physical size or expansion velocity.  
Here the most recent compilation based on {\it Gaia} DR3 data is used \citep{Bailer2021}.  
Distances to novae have been critically tested by \citet{Schaefer2022}, which has allowed us to extend the sample of distances to novae whose parallaxes are not available or deemed unreliable, and refine those with available parallaxes.  
A comparison of distances to the sample of nova remnants with available images is presented in Appendix~\ref{app.dist}.

There is a number of notorious sources catalogued as nova remnants that can be missed in Table~\ref{tab:nov}.  
Some of them are among the largest nova remnants, including some ancient nova shells and nova super remnants candidates, whereas the nova shell nature is disputed for others.  
The properties of these sources are discussed in Appendix~\ref{app1} and they are listed in Table~\ref{tab:other_nova}. The reader is referred to the excellent compilation of {\it bona fide} and candidate ancient novae provided by \citet{Tappert+2020}.

\subsection{General properties}

The nova spatial distribution in the Galaxy has been recently discussed in length by \citet{Schaefer2022}.  
The Aitoff projection in Fig.~\ref{fig:aitoff} shows that most novae are projected close to the Galactic Plane and along the direction of the Galactic Center, suggesting that they are associated with the general stellar distribution in the Galaxy. 
The number of novae along the direction of the Galactic Center could be even larger, as extinction may hide a significant fraction of them, and crowding avoids their detection.    
Moreover, there is a noticeable lack of novae behind the Galactic Center in the top-left panel of Fig.~\ref{fig:distrib}, most likely suggesting that the high extinction along this line of sight hampers their detection at large distances.

It is interesting to compare the spatial distributions of novae and those of resolved nova remnants.  
Nova remnants can be found up to large distances but, compared to the general population of novae, there is a notorious deficit of nova remnants at distances larger than $\approx$5000 pc (top-right panel of Fig.~\ref{fig:distrib}) and at Galactic latitude $>$30 degrees (bottom-right panel of Fig.~\ref{fig:distrib}), i.e., the location of the Galactic Center and its Bulge.  
The large distance, high extinction, and stellar crowding along the direction of the Galactic Center all play together to produce a bias against the detection of the nebular emission of bulge nova remnants. 
 
Comparing the spatial distributions of detected nova remnants with those undetected also proves to be informative.
Both resolved (Table~\ref{tab:nov}) and undetected (Table~\ref{tab:nov_st}) nova remnants are similarly distributed in Galactic latitude and longitude coordinates (bottom panels of Fig.~\ref{fig:distrib}), but resolved nova remnants have a higher chance to be detected at distances below $\leq$2500 pc and above $\geq$5000 pc (top-right panel of Fig.~\ref{fig:distrib}).  
The proximity of nova remnants closer than 2500 pc certainly helps in detecting them, as extinction is smaller and their observed H$\alpha$ surface brightness is higher.  
At large distances above 5000 pc, however, the higher occurrence rate of nova remnants compared to those undetected cannot be explained on the basis of distance.

\subsection{Images}\label{images}

\begin{figure*} 
\centering 
\includegraphics[width=0.95\linewidth]{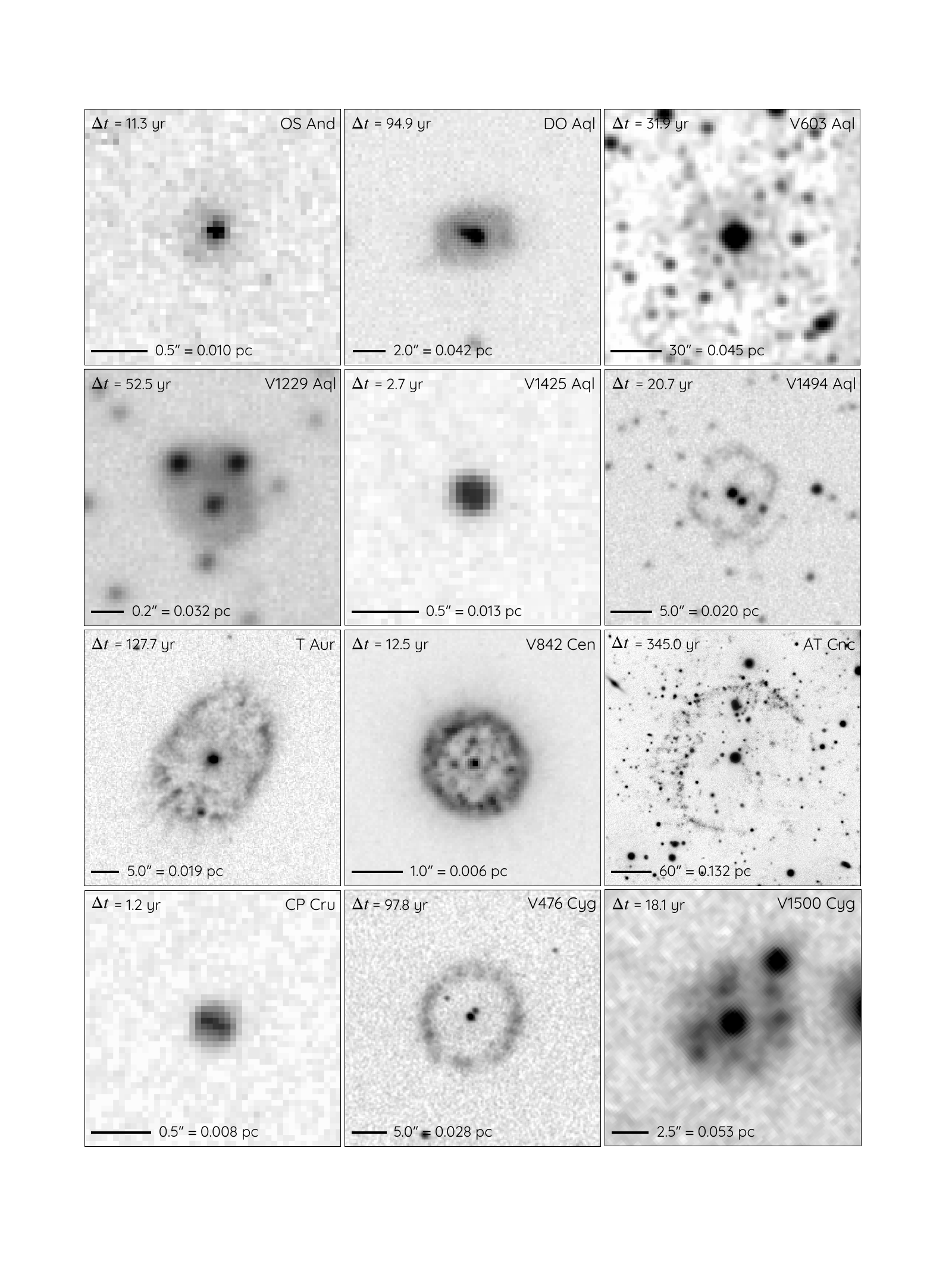}
\caption{
Optical images of the nova remnants OS\,And, DO\,Aql, V603\,Aql (top row), V1229\,Aql, V1425\,Aql, V1494\,Aql (second raw), T\,Aur, V842\,Cen, AT\,Cnc (third row), CP\,Cru, V476\,Cyg and V1500\,Cyg (bottom row).  
All images show emission of the H$\alpha$ and/or [N~{\sc ii}] emission lines as obtained using narrow-band filters, but that of V603\,Aql obtained through a long-pass filter. 
The time of the image since the nova outburst is labelled, as well as the angular scale and its corresponding linear scale at the distance listed in Tab~\ref{tab:nov}.
North is up, East to the left.
} 
\label{1.img} 
\end{figure*}

\begin{figure*} 
\centering 
\includegraphics[width=0.95\linewidth]{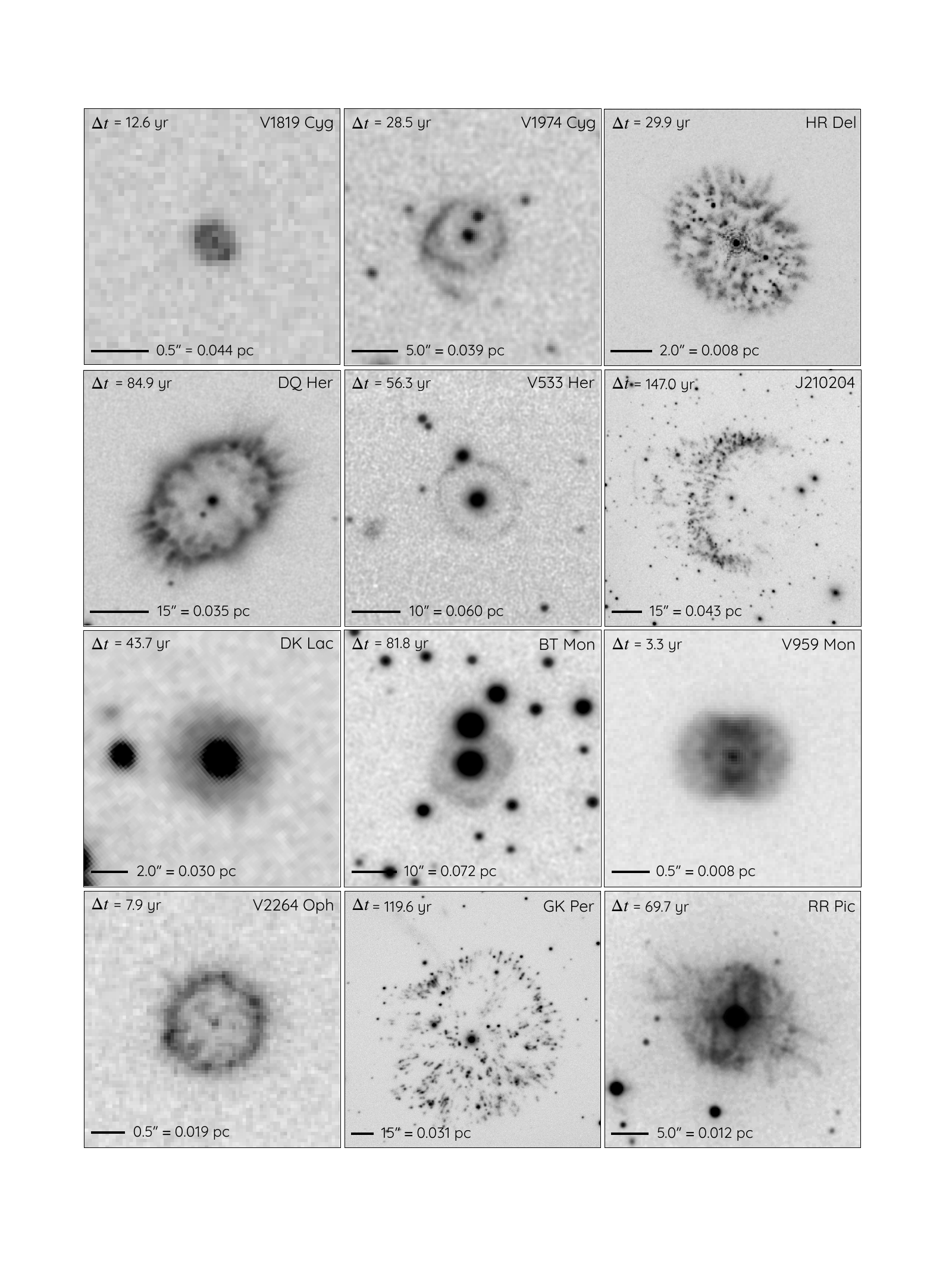}
\caption{Same as Fig.~\ref{1.img} for the nova remnants. Optical images of the nova remnants V1819\,Cyg, V1974\,Cyg, HR\,Del (top row), DQ\,Her, V533\,Her, J210204 (second raw), DK\,Lac, BT\,Mon, V959\,Mon (third row), V2264\,Oph, GK\,Per and RR\,Pic (bottom row). North is up, East to the left.
} 
\label{2.img} 
\end{figure*}

\begin{figure*} 
\centering 
\includegraphics[width=0.95\linewidth]{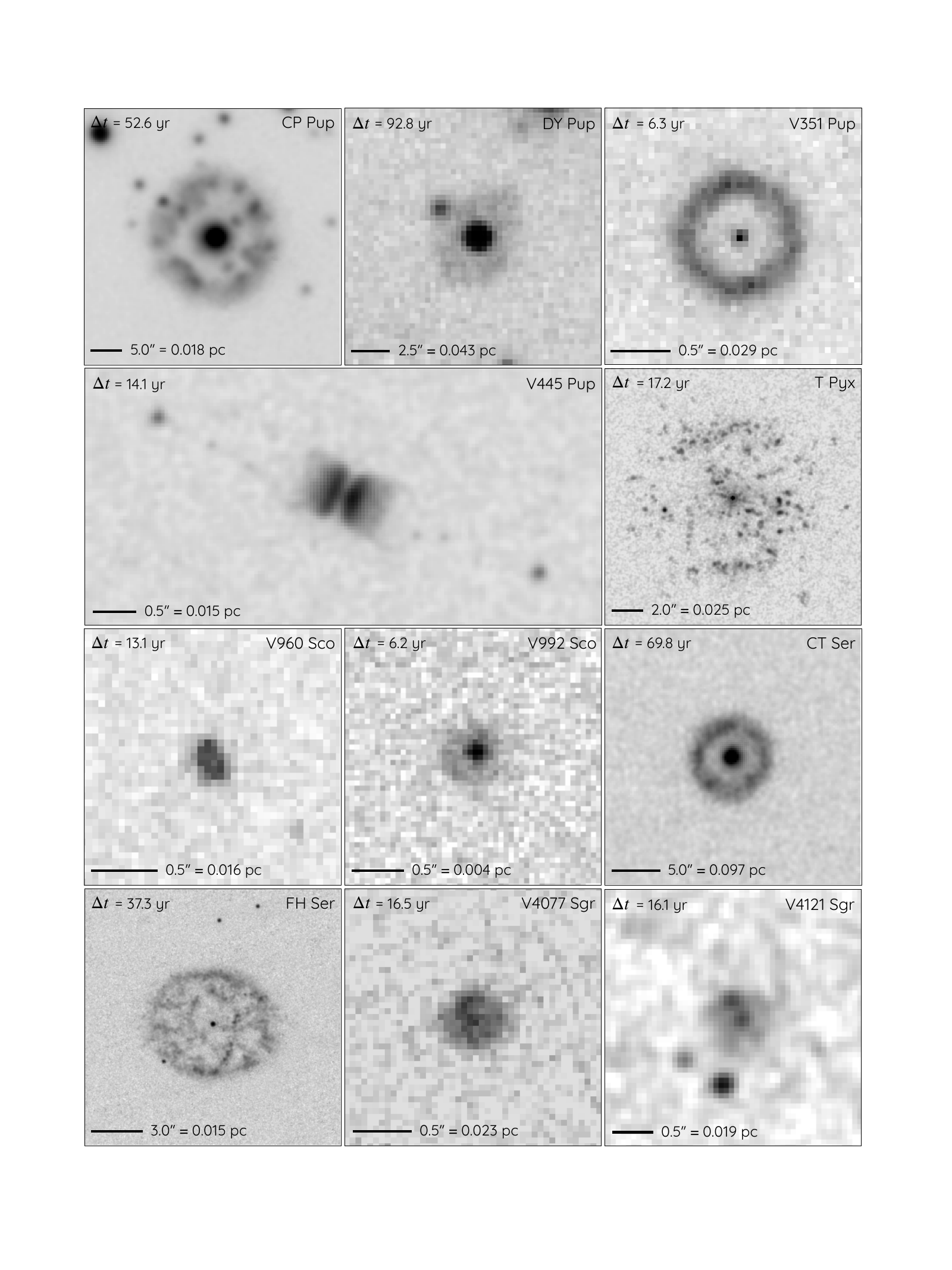}
\caption{Same as Fig.~\ref{1.img} for the nova remnants. Optical images of the nova remnants CP\,Pup, DY\,Pup, V351\,Pup (top row), V445\,Pup, and T\,Pyx (second raw), V960\,Sco, V992\,Sco, CT\,Ser (third row), FH\,Ser, V4077\,Sgr and V4121\,Sgr (bottom row). V445\,Pup was obtained through of the HST F502N filter. North is up, East to the left.
} 
\label{3.img} 
\end{figure*}

\begin{figure*} 
\centering 
\includegraphics[width=0.95\linewidth]{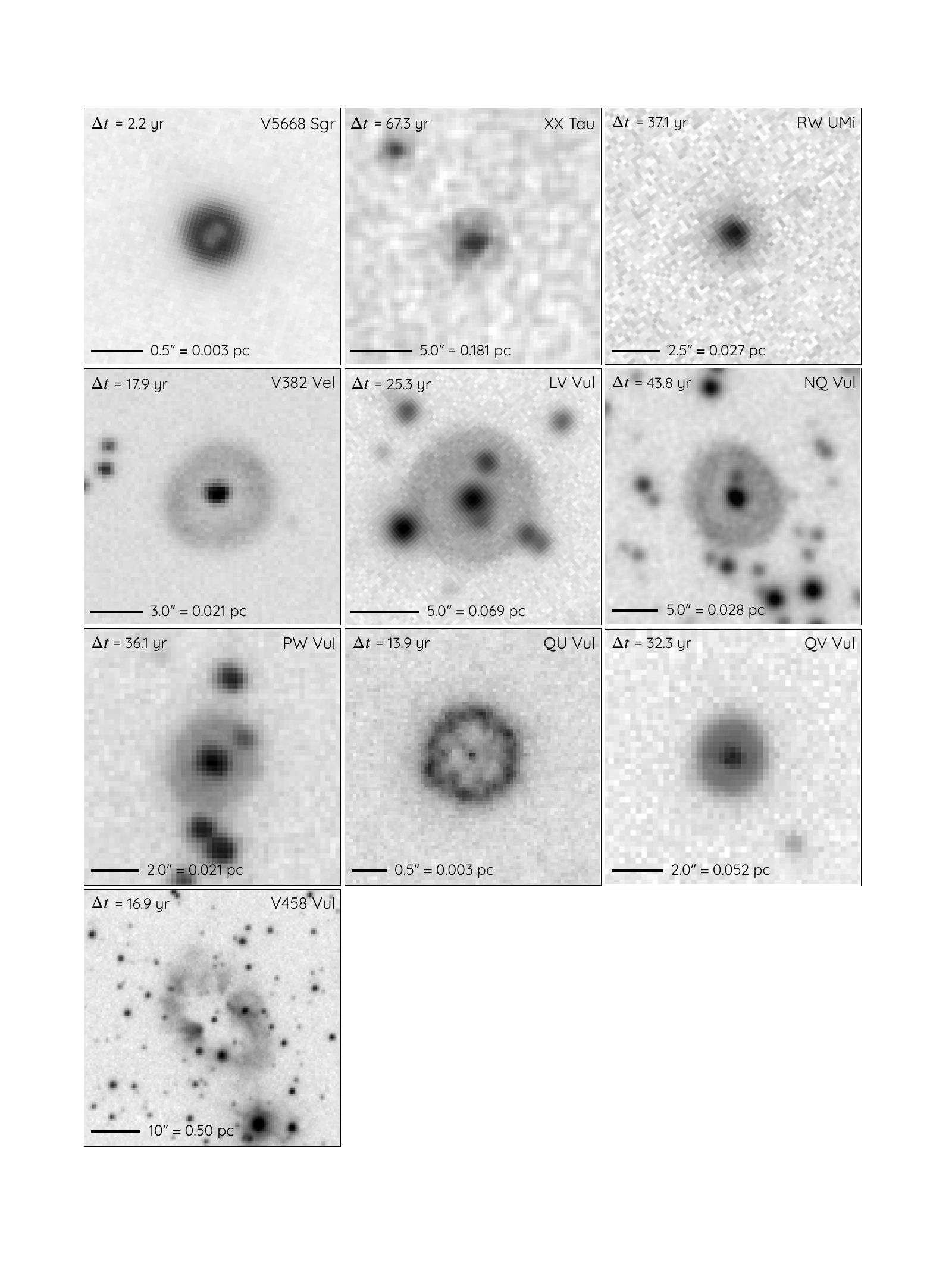}
\caption{Same as Fig.~\ref{1.img} for the nova remnants. Optical images of the nova remnants V5668\,Sgr, XX\,Tau, RW\,UMi (top row), V382\,Vel, LV\,Vul, NQ\,Vul (second raw), PW\,Vul, QU\,Vul, QV\,Vul (third row), and V458\,Vul (bottom row). North is up, East to the left.} 
\label{4.img} 
\end{figure*}

The optical images of the resolved nova remnants are presented in Figs.~\ref{1.img}, \ref{2.img}, \ref{3.img} and \ref{4.img}. 
The set obtained at NOT using the ALhambra Faint Object Spectrograph and Camera (ALFOSC) corresponds to our own observing programs of nova remnants on several epochs.
The rest corresponds to archival images obtained through nebular filters that register the emission of the H$\alpha$ or H$\alpha$+[N~{\sc ii}] emission lines, with the only exception of the images of V603\,Aql, obtained using a long-pass red filter, and V445\,Pup, obtained using the \emph{HST} WFPC3 F502N [O~{\sc iii}] filter.

The images are presented using a log-inverted grey scale. 
In a number of cases, multi-epoch images were available \citep[e.g., DQ\,Her,][]{Santa+2020}.  
Then the best available image was selected for its presentation here.  
An investigation of the angular expansion of nova remnants with multi-epoch observations is in course (Santamar\'ia et al., in preparation). 
Comments on a few iconic nova remnants are given in Appendix~\ref{app2}.

\section{Results}

\subsection{Morphology}\label{morph}

The images presented in Fig.~\ref{1.img}, \ref{2.img}, \ref{3.img}, and \ref{4.img} have been used to assess a morphological classification for the resolved nova remnants in Table~\ref{tab:nov}, \ref{tab:nov_st}, and \ref{tab:other_nova}.   
The ERBIAS scheme used for planetary nebulae  \citep[PNe;][]{Parker+2006} is adopted for the morphology classification of the main shape of nova remnants. 
The upper-case ERBIAS stands for (E)lliptical, (R)ound, (B)ipolar (or hourglass shaped), (I)rregular, (A)symmetric, and (S)tellar major morphological classes. 
A nova remnant is classed as ``E'' instead of ``R'' when it has an oval shape and its aspect ratio exceeds the arbitrary value of 1.05, i.e., its major axis is at least 5 per cent larger than its minor axis. 
The ``S'' class obviously applies to the novae without resolved nebular emission listed in Tab~\ref{tab:nov_st}.  
The FWHM of the point-spread function (PSF) of the central star of the novae in this table has been compared with that of stars in the field of view (FWHM$_\mathrm{inst}$).  
If found to be larger, implying the presence of unresolved nebular emission, these have been classified as (G)aussian.

Additional small-scale features or secondary morphological attributes will be described by the lower-case ``scxftme'' letters.  
These stand for (s)mooth, (c)lumpy or mi(x)ed (smooth and clumpy) surface brightness, (f)ilaments, (t)ails associated with cometary knots, (m)ultiple shells, and (e)nhancement of the surface brightness enhancement along the minor axis.  
The latter is suggestive of a brighter equatorial structure.

The morphological classification of the nova remnants is listed in column \#11 of Table~\ref{tab:nov} and column \#5 of Table~\ref{tab:other_nova}. 
It is worth noting that the optical filters used to register the images of the nova remnants presented here may include emission from both the H$\alpha$ and [N~{\sc ii}] $\lambda\lambda$6548,6584 \AA\ emission lines, depending on the filter band-pass width and expansion velocity of the nova remnant.  
One example is FH\,Ser, with morphology Ecef according to its \emph{HST} WFPC2 F656N image, which is revealed by integral field spectroscopic observations to consist of an H$\alpha$ bright ellipsoidal shell and a [N~{\sc ii}] clumpy equatorial ring \citep{Guerrero+2025b}.  
The situation is even more extreme in J210204, with morphology Ic, whose brightest clumps are actually dominated by [N~{\sc ii}] emission, which is about 25 times brighter than H$\alpha$ \citep{Guerrero+2018}.

\begin{figure}
\centering
\includegraphics[width=1.0\linewidth]{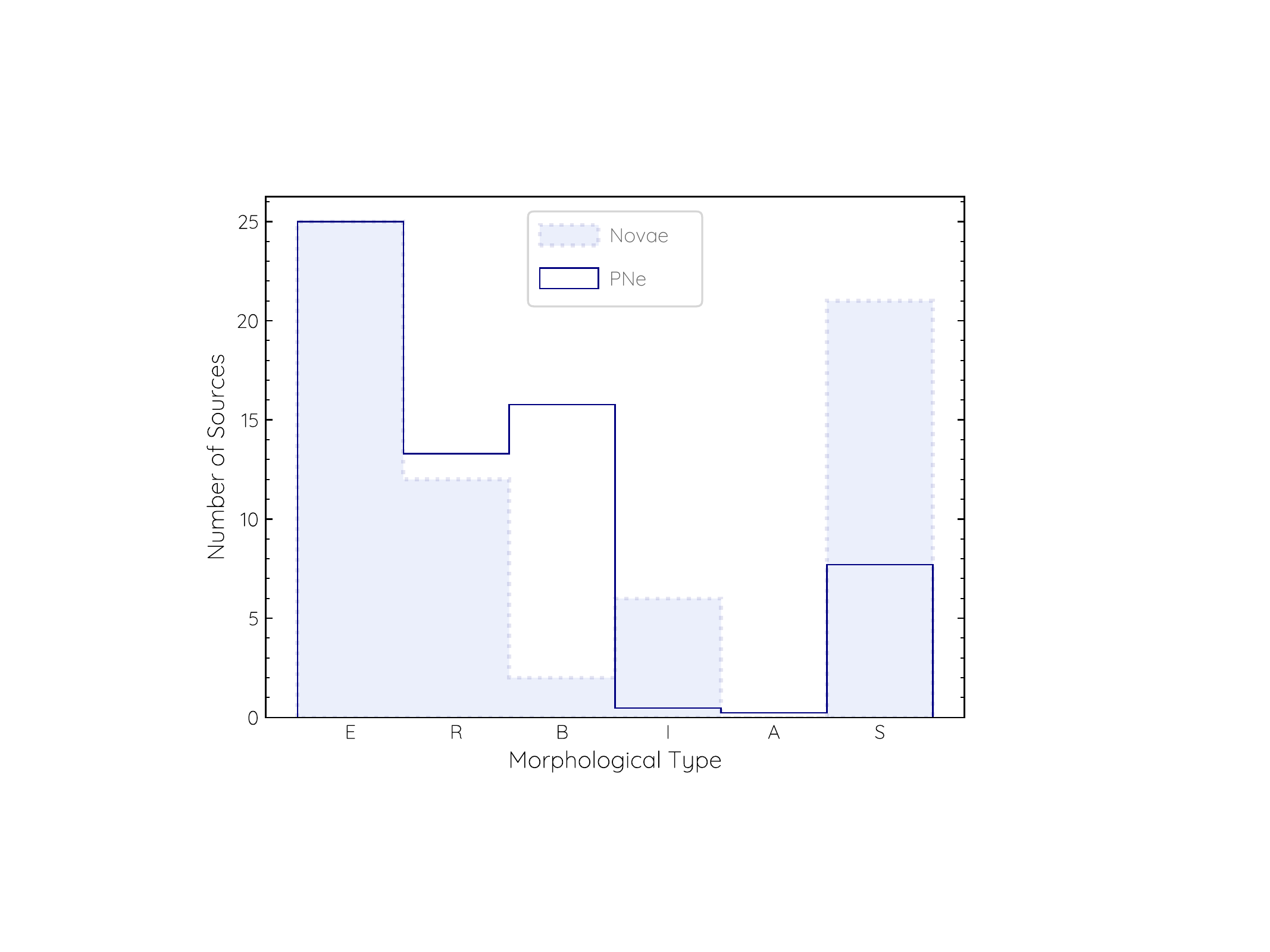}
\caption{
Morphological type distributions of nova remnants and PNe according to the classification scheme for PNe \citep{Parker+2006}. 
The morphological distribution of PNe is derived from the data provided by their table~2. 
}
\label{fig:histo}
\end{figure}

Most nova remnants listed in Table~\ref{tab:nov} are either elliptical or round, with very few bipolar nebulae.  
This is at odds with PNe, as shown in Fig.~\ref{fig:histo}, with a significant fraction of PNe having a bipolar morphology \citep{Parker+2006}.  
Given the statistical meaning of the sample of nova remnants, this difference can not be attributed to orientation effects.  
It most likely reflects that the time-scale of the interaction of the nova ejecta with the accretion disk and binary companion is too short to have deep effects on the nova shaping.  
On the contrary, the time-scales of the interactions between stellar winds of different geometry in PNe are much longer, resulting in more noticeable effects on the nebular shaping.  
It is worth noting that nova remnants and PNe have quite similar ratios between elliptical and round morphologies, with about two elliptical sources for each round one.

If we exclude RS\,Car and GQ\,Mus with the largest distances, the G and S unresolved nova remnants have similar age and distance distributions, with mean ages around 60 and 66 yr and mean distances around 3800 and 3200 pc, respectively.  
Note that four out of the five oldest nova remnants in Table~\ref{tab:nov_st}, namely RS\,Car, AR\,Cir, V841\,Oph, and CN\,Vel,  have an S morphology, whereas the detection of diffuse emission in the fifth one, IL\,Nor, is questionable.  
These old nova remnants may have already dispersed into the surrounding ISM below detection limits.

As for the small-scale features or secondary morphological attributes, it is worth noting that the capability to detect knots or clumps might be affected by the spatial resolution \citep[see the effects of spatial resolution on the apparent clumpiness of the nova remnant V5668\,Sgr described by][]{Diaz+2018}, very particularly when the spatial resolution is comparable to the angular radius of the source. 
Otherwise, the detection of filaments (f) and cometary knots with tails (t) is also associated with clumps, which seems to confirm that a minimal spatial resolution compared to the object's angular size is required to detect these small-scale features. 
Moreover, there are very few sources with multiple shell (m) or equatorial enhancement (e) features.  
The latter is to be connected with the low occurrence of bipolar morphologies.

\subsection{Angular and physical radius of nova remnants}

The angular radii of resolved nova remnants along the minor ($\theta_\mathrm{min}$) and major ($\theta_\mathrm{max}$) axes are listed in column \#12 of Table~\ref{tab:nov}.
These have been measured as the angular extent at an intensity 5-$\sigma$ over the background level of the image.  
Nova remnants with a G morphology are indicative of nebular yet unresolved emission.  
The angular radius of these in column \#11 of Table~\ref{tab:nov_st} is derived by comparing the observed FWHM of the nova (FWHM$_\mathrm{obs}$) to the instrumental FWHM (FWHM$_\mathrm{inst}$) given by stellar sources in the field of view to obtain the intrinsic instrumental-corrected FWHM (FWHM$_\mathrm{o}$) as
\begin{equation}
    \mathrm{FWHM}_{\rm o}^2 = \mathrm{FWHM_{\rm obs}}^2-\mathrm{FWHM}_{\rm inst}^2, 
\end{equation}
\noindent 
adopting then the relationship 
\begin{equation}
    \theta = \frac{\mathrm{FWHM}_{\rm o}}{2 \sqrt{\ln 2}},
    \label{eq.G_rad}
\end{equation}
\noindent 
described by \citet{Santamaria+2022}. 
Column \#9 of Table~\ref{tab:nov_st} provides the values of FWHM$_\mathrm{inst}$, which,  
for unresolved S nova remnants sets 
an upper limit of the angular extent of any possible nebular emission.

The physical radius of the resolved nova remnants in Table~\ref{tab:nov}, derived from their distance and average angular radius 
\begin{equation}
    r = d \times \theta,
\end{equation}
is plotted vs.\ their age at the time the image was acquired ($\tau$) and distance ($d$) in Figs.~\ref{fig:s2age} and \ref{fig:s2d}, respectively\footnote{Nova remnants with a G morphology are not included in these plots, since their angular radius is derived in a different way than those of resolved nova remnants.}.  
In general, there is an increase in the physical size of nova remnants with their age, as expected from their expansion, which can be described by the linear relationship,
\begin{equation}
\left( \frac{r}{\mathrm{pc}} \right)= (0.000725 \pm 125) \left( \frac{\tau}{\mathrm{yr}} \right)
\label{eq.svst}
\end{equation}
or alternatively
\begin{equation}
\left( \frac{r}{\mathrm{km}} \right) = (725 \pm 125) \left( \frac{\tau}{\mathrm{s}}\right),
\label{eq.svst2}
\end{equation}
implying an average expansion velocity of $725\pm125$ km~s$^{-1}$ for the sample.

The physical radius of nova remnants is generally smaller than 0.10 pc, but about twice for the nova remnant AT\,Cnc, and a few times this value for the ancient nova shells V1315\,Aql (0.52 pc), Z\,Cam (0.9 pc for the inner shell and 2.5 pc for the outer shell), V1316\,Cyg (0.29 pc along the minor axis and 0.41 along the major axis), and RX\,Pup (0.17 pc for the inner shell, but 0.93 pc for the outer shell).  
It is worth noting that the physical sizes of AT\,Cnc, V1315\,Aql and Z\,Cam are in agreement with their age according to the relationship given by Eqs.~(\ref{eq.svst}) and (\ref{eq.svst2}), whereas that of 
RX\,Pup is several times smaller than expected.

\subsection{Age of nova remnants}

\begin{figure}
\centering
\includegraphics[width=1.0\linewidth]{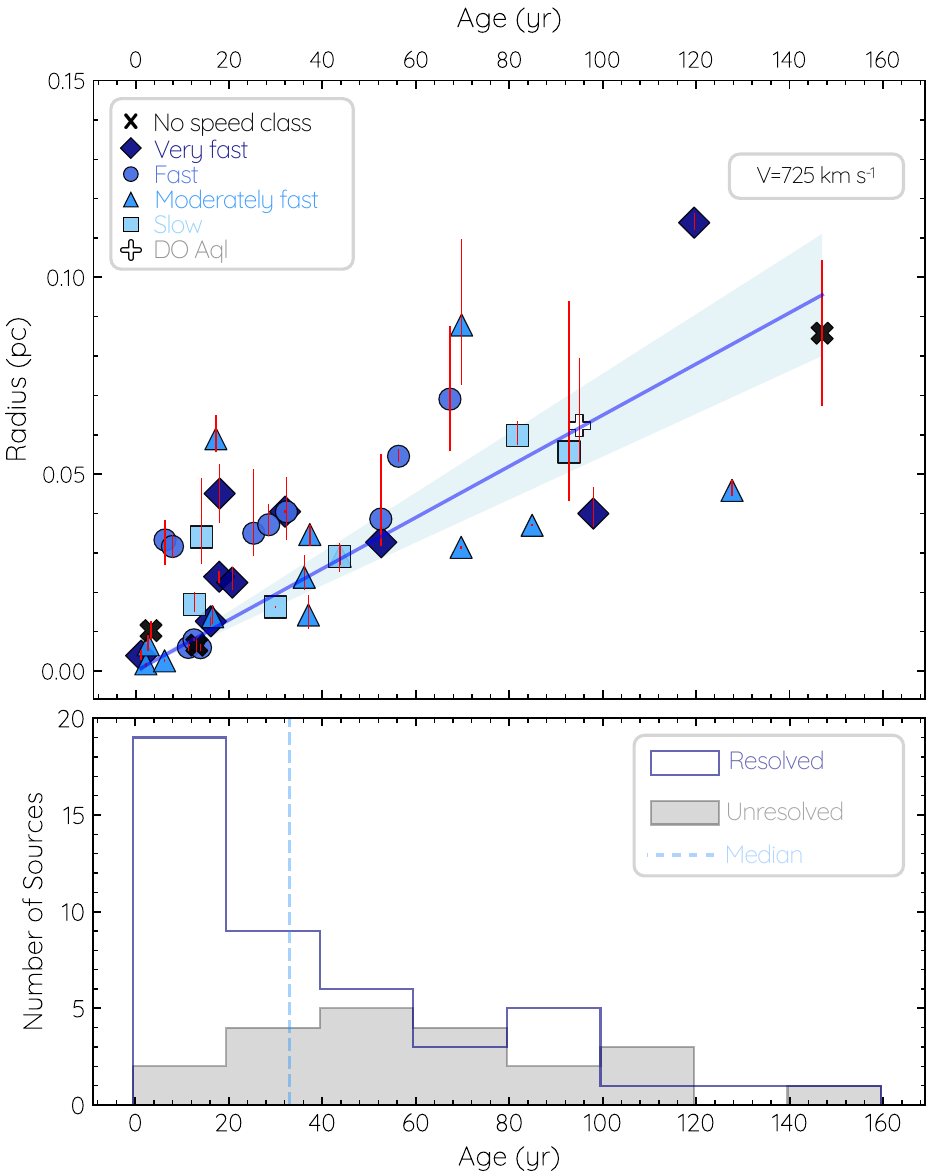}
\caption{
Distribution of physical radius ($d\times\theta$) with age ($\tau$) of resolved nova remnants.  
The averaged angular radii and age correspond to the time when the images presented here were acquired. 
The solid blue line is a fit to data points and the blue-shaded area represent the 95 per cent confidence band.  
AT\,Cnc, with a radius of 0.17 pc and an age in the range 325-345 yr is out of the limits of this figure, but it follows the linear relation-ship.  
The inset shows the age distributions of resolved (red histogram) and unresolved/undetected (grey histogram) nova remnants.  
The vertical line corresponds to the median age of 33.1 yr of resolved nova remnants. 
}
\label{fig:s2age}
\end{figure}

The bottom panel in Fig~\ref{fig:s2age} shows the age distribution of resolved and unresolved nova remnants. 
The age distribution of resolved nova remnants follows a decay almost exponential, with a median age of 33.2 yr (dashed vertical line in the plot) and over 75 per cent of them with ages younger than 60 yr.  
On the other hand, there is a notorious lack of stellar/undetected nova remnants among the youngest sources, with a broader distribution around a median value of $\simeq$60 yr. 
This difference is aggravated by the smaller average distance of stellar/undetected nova remnants, which should have improved their detection occurrence.  
The comparison of both distributions shows that $\simeq$80\% of nova remnants younger than 60 yr are detected, but only $\simeq$50\% of older nova remnants.

The age distribution of resolved nova remnants probes that their detection rate decreases with age, which can be expected as the nova remnant expands and dims to fall below the detection limit of available images \citep{Tappert+2020}. 
The age of 150 yr seems to place an upper limit for the visibility time lapse of nova remnants. 
Exceptions are the few examples of ancient nova shells V1315\,Aql ($\tau = 500 - 1200$ yr), Z\,Cam ($\tau = 2700$ yr), and RX\,Pup ($\tau = 1300$ yr) described in Appendix~\ref{app1} (the other ancient nova shell candidate V1363\,Cyg is not confirmed, neither is  there an estimate of its age). 
The other exception is AT\,Cnc, whose age is in the range from 325 to 365 yr.

\begin{figure}
\centering
\includegraphics[width=1.0\linewidth]{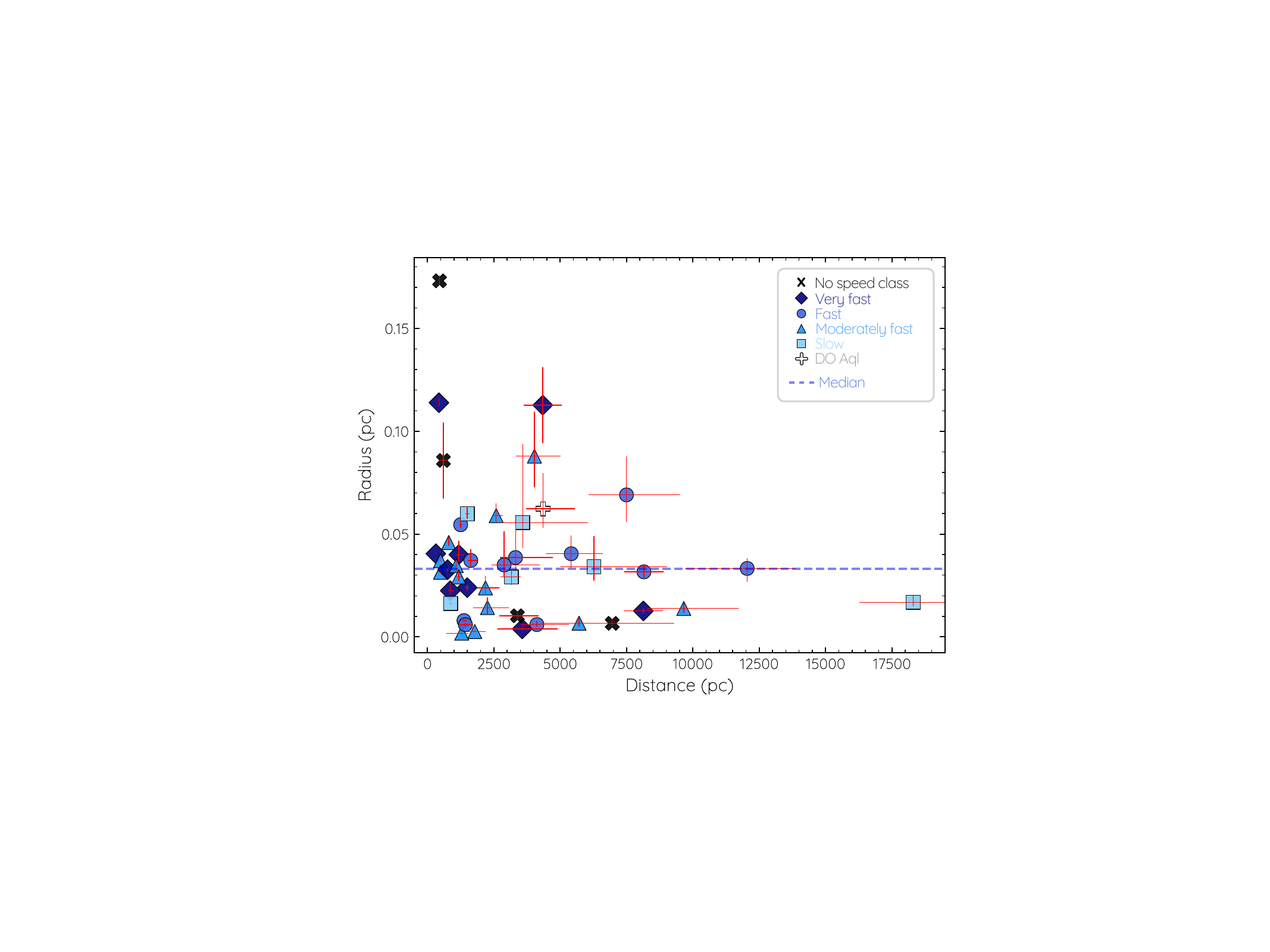}
\caption{
Distribution of physical radius ($d\times\theta$) with distance ($d$) of resolved nova remnants. 
The median value of the physical radius, $\approx$0.03 pc, is shown by the horizontal blue line.
}
\label{fig:s2d}
\end{figure}

Fig.~\ref{fig:s2d} shows that closer nova remnants seem to have larger physical radii (and thus older ages), which implies there is a bias against the detection of old nova remnants at large distances.  
Indeed only a few nova remnants are detected beyond 8 kpc and all of them are smaller than 0.04 pc in radius.  
This bias can be attributed to the decline in surface brightness with age \citep{Tappert+2020} and the general increase of extinction with distance.

\subsection{Expansion velocity of resolved nova remnants}

\begin{figure}
\centering
\includegraphics[width=1.0\linewidth]{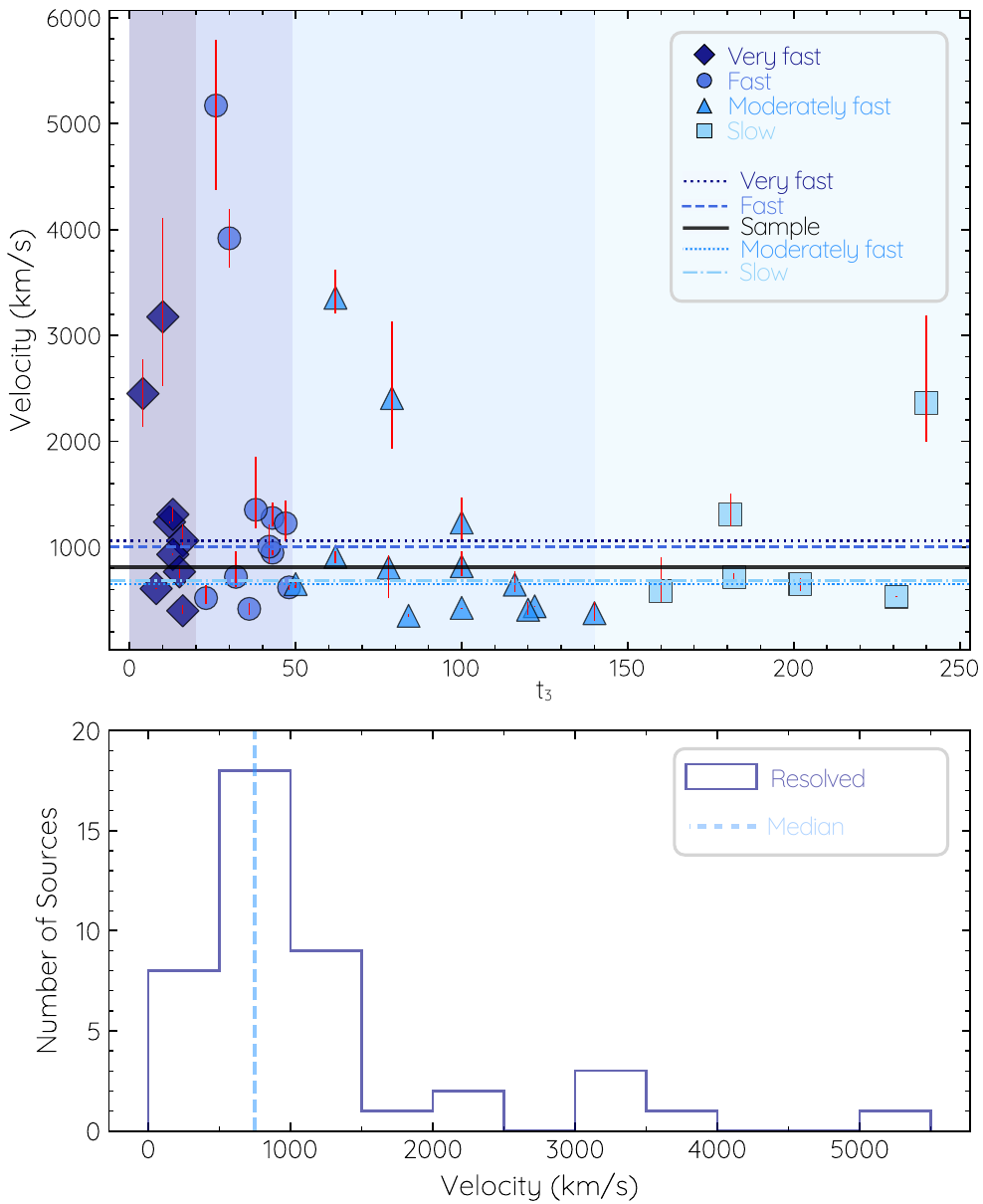}
\caption{ 
Distribution of the average expansion velocity ($d\times\theta/\tau$) and decline time \citep[$t_3$,][]{Gaposchkin1957} of the resolved nova remnants. The horizontal dashed black line corresponds to a median value of 810 km~s$^{-1}$. Data points are coloured using a gradient proportional to the error magnitude, following the 'Blues' colormap. 
The plot highlights four regions along the $x$-axis, which indicates the speed class  ($t_{3}$). 
We note that the slow nova DO\,Aql, with its extremely long $t_3$ of 900 days, is not displayed in this figure.  
Its mean expansion velocity of 670 km~s$^{-1}$ is consistent with the average value of slow novae. 
}
\label{fig:v2t3}
\end{figure}

The mean expansion velocity of the nova remnants listed in Table~\ref{tab:nov} can be derived from their distance, average angular radius and image age as 
\begin{equation}
    V = \frac{d \times \theta}{\tau}.  
\end{equation}
These are plotted vs.\ the decline time ($t_3$) in the top panel of Fig.~\ref{fig:v2t3}.  
The average and median expansion velocity of this sample of nova remnants are 865 and 930 km~s$^{-1}$, respectively, as shown by the horizontal dashed line in Fig.~\ref{fig:v2t3}\footnote{
This figure differs from the expansion velocity of 725 km~s$^{-1}$ provided in the previous section as they have been derived using different methods (linear fit vs.\ average) and cast from different samples. }.

The average mean expansion velocity for each speed class is also illustrated in the top panel of Fig.~\ref{fig:v2t3} and listed in Table~\ref{tab:v2vclass}. 
In general, there is a decline of the expansion velocity with time, from very fast, to fast, and to moderately fast and slow objects.  
For the latter moderately fast and slow\footnote{
The apparent discrepant higher expansion velocity of slow novae compared to moderately fast novae would be reversed if RW\,UMi (mean expansion velocity of 385 km~s$^{-1}$), a moderately fast nova with a $t_3$ value of 140 days at the lower limit of slow novae, were instead considered a slow nova.}, 
the average mean expansion velocity stalls around a mean value of 650--680 km~s$^{-1}$.

\begin{table}
\centering
\setlength{\columnwidth}{1.0\columnwidth}
\setlength{\tabcolsep}{1.0\tabcolsep}
\caption{Average mean expansion velocity of nova remnants according to their speed class.}
\label{tab:v2vclass}
\begin{tabular}{lrcc}
\hline
\hline

\multicolumn{1}{l}{Nova Speed Class} &
\multicolumn{1}{c}{$t_3$} &
\multicolumn{1}{c}{Average mean $V_\mathrm{exp}$} &
\multicolumn{1}{c}{Average} \\

\multicolumn{1}{c}{} &
\multicolumn{1}{c}{(day)} &
\multicolumn{1}{c}{(km~s$^{-1}$)} & 
\multicolumn{1}{c}{aspect ratio} \\
\hline
All             & $\dots$~~~~~ &  810 & 1.15 \\
Slow            &  $141-264$ &  685 & 1.13 \\
Moderately fast &   $50-140$ &  650 & 1.11 \\
Fast            &  $21-49$~~ & 1000 & 1.16 \\
Very fast       & $<$20~~~~~ & 1060 & 1.24 \\
\hline
\end{tabular}
\vspace{0.15cm}
\end{table}

The histogram at the bottom panel of this figure shows the frequency of occurrence of expansion velocity in this sample of nova remnants.  
The distribution peaks at the bin from 500 to 1000 km~s$^{-1}$, in agreement with the linear fit to Fig.~\ref{fig:s2age}.  
The distribution shows a noticeable high-velocity tail, with V351\,Pup being the highest velocity nova remnant in this sample, although it should be considered with caution given its large distance ($12.0^{+1.8}_{-2.3}$ kpc).

\subsection{Aspect ratio of shell-like nova remnants}

\begin{figure*}
\centering
\includegraphics[width=1.0\linewidth]{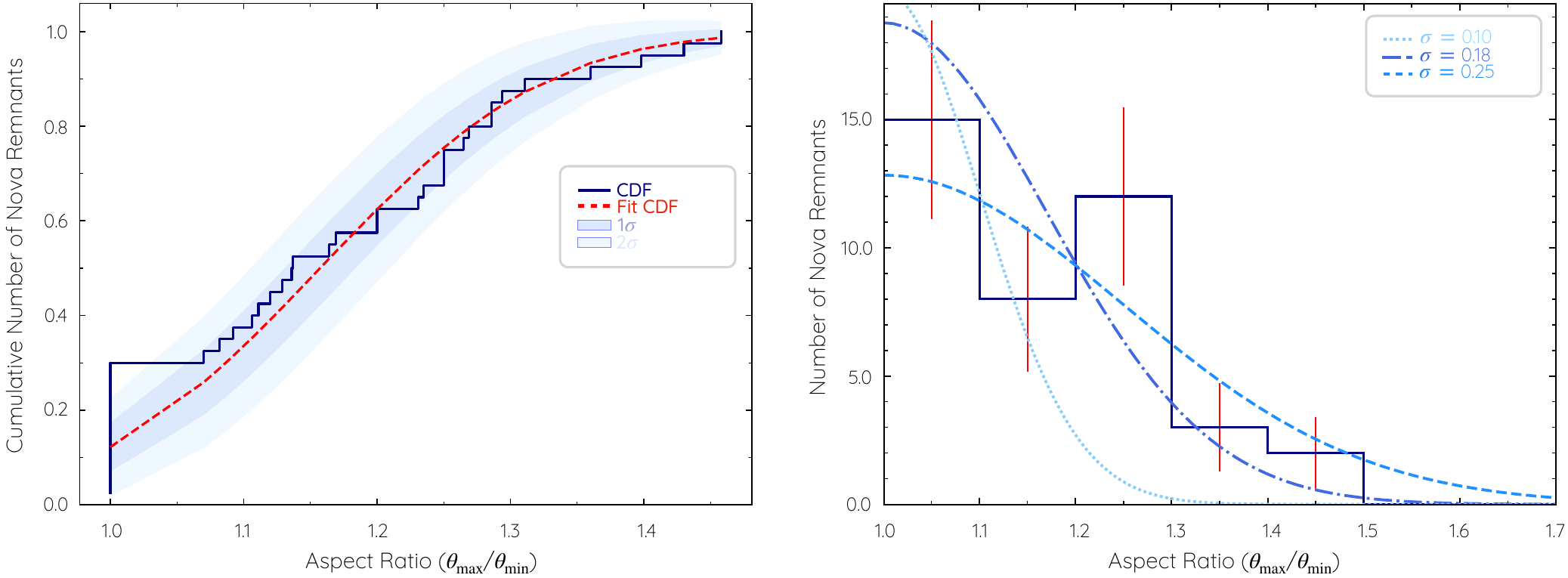}
\caption{
(left) Cumulative distribution function (CDF) is shown as a step plot (solid-line) and (right) histogram (solid-line histograms) of the aspect ratio ($\theta_\mathrm{max}/\theta_\mathrm{min}$) of nova remnants with a shell-like elliptical or round morphology. 
In the cumulative distribution plot in the left panel, the red dashed line represents the CDF of the normal distribution fitted to the observed data. The Kolmogorov-Smirnov (KS) test yielded a statistic of 0.18 with a p-value of 0.14, indicating a reasonable fit of the normal distribution to the data. The blue-shaded areas represent the 1-$\sigma$ and 2-$\sigma$ confidence bands. These Gaussian distributions, normalized to the number of nova remnants, are shown with the same line styles and colours in the histogram plot in the right panel. 
}
\label{fig:axial_rat}
\end{figure*}

The most frequent morphology among the resolved nova remnants listed in Table~\ref{tab:nov} is elliptical (56\%), with another significant fraction having a round morphology (27\%).  
It is very likely that projection effects make some elliptical nova remnant to appear round. 
The cumulative distribution of the observed aspect ratio ($\theta_\mathrm{max}/\theta_\mathrm{min}$) of nova remnants with a shell-like elliptical or round morphology is shown in Fig.~\ref{fig:axial_rat}-left. 
About 30\% of the shell-like nova remnants in the sample have round morphology, that has been defined as those with $\theta_\mathrm{max}/\theta_\mathrm{min} < 1.05$.  
Then the cumulative distribution increases steadily up to values of $\theta_\mathrm{max}/\theta_\mathrm{min} \approx 1.3$, when it flattens, with not a single source with $\theta_\mathrm{max}/\theta_\mathrm{min}$ above 1.45.

The cumulative distribution of observed $\theta_\mathrm{max}/\theta_\mathrm{min}$ contains information on their true, deprojected distribution.  
The cumulative distribution of observed $\theta_\mathrm{max}/\theta_\mathrm{min}$ is overplotted in Fig.~\ref{fig:axial_rat}-left with the expected cumulative distributions for the true aspect ratio assuming it follows a Gaussian distribution starting from a round shape ($\theta_\mathrm{max}/\theta_\mathrm{min}  = 1.0$) with a given value of its width $\sigma$.  
The Kolmogorov-Smirnov (KS) test indicates that a Gaussian distribution with $\sigma$ of $0.18\pm0.07$ fits reasonably well the observed cumulative distribution of aspect ratios. Accordingly, about 68\% of nova remnants have aspect ratios smaller than 1.25 (1-$\sigma$), with less than 5\% having aspect ratios greater than 1.32 (2-$\sigma$).

These distributions of aspect ratios are presented in the right panel of Fig.~\ref{fig:axial_rat}, compared to the observed distribution of aspect ratios.  
Although the broadest distributions predict the presence of nova remnants with values of the observed aspect ratio above 1.5, these are not detected.  
This is most certainly due to projection effects, which play against the detection of sources with the major axis close to the plane of the sky, and to the limited number of sources in the sample.

\begin{figure}
\centering
\includegraphics[width=1.0\linewidth]{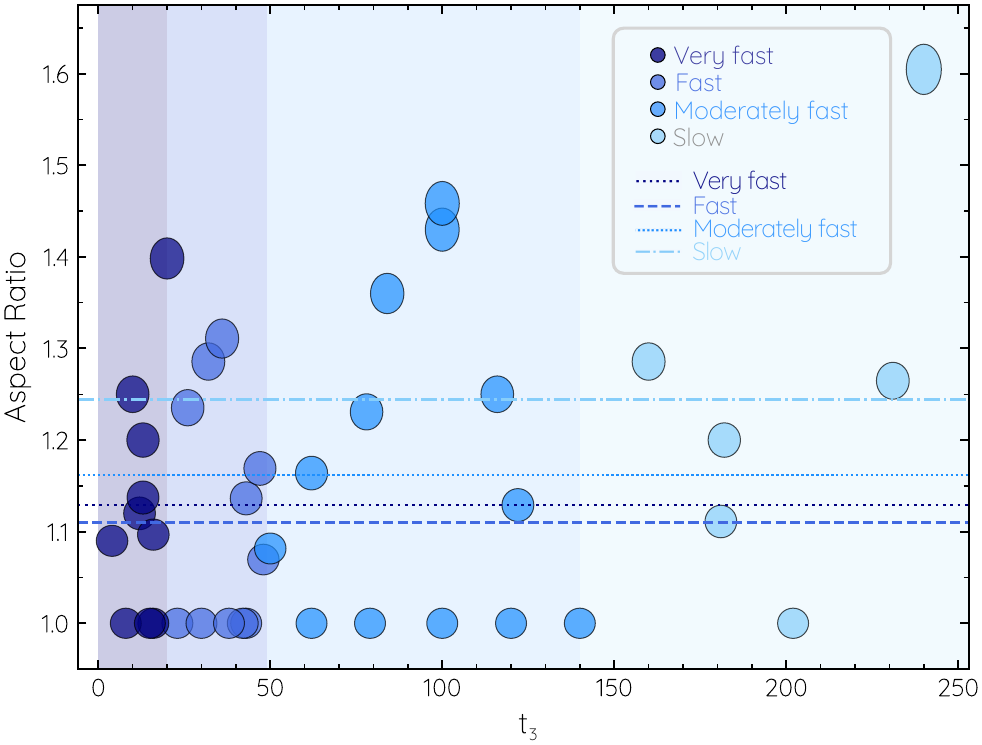}
\caption{
Distribution of the aspect ratio with the decline time $t_3$, the 'Blues' colormap indicate speed classes. As the aspect ratio increases, the ellipticity of each sample point correspondingly intensifies. The average mean expansion velocities of nova remnants, categorized by speed class, are depicted on each horizontal line, with the corresponding estimated values provided in the Tab.~\ref{tab:v2vclass}.
}
\label{fig:axrat2t3}
\end{figure}

\begin{figure}
\centering
\includegraphics[width=1.0\linewidth]{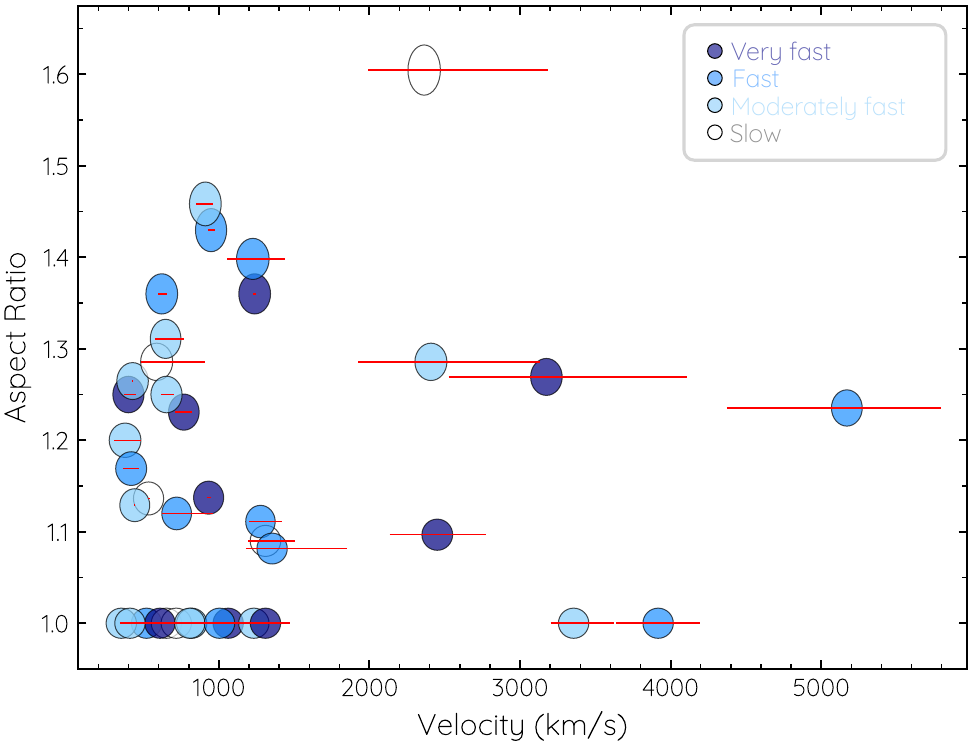}
\caption{
Comparison of the observed aspect ratio with the expansion velocities of nova remnants. The circular 'Blues' dots represent the circular nova shells with ‘spherical’ morphologies, whereas the elliptical points correspond to ellipsoidal shells. All dots are categorized by their speed class.
}
\label{fig:velax}
\end{figure}

The ejecta of nova remnants with longer decline times $t_3$ and thus slower expansion velocities $v_{\rm exp}$ are expected to experience more prolonged interactions with the accretion disk and binary companion.  
The nova remnant shaping will be stronger in these cases, and thus an anti-correlation between aspect ratio and $t_3$ and $v_{\rm exp}$ can result.  
These are investigated in Figs.~\ref{fig:axrat2t3} and \ref{fig:velax}.  
There is not a clear correlation, however.  
The average aspect ratios of the different speed classes of nova remnants are listed in the last column of Tab.~\ref{tab:v2vclass}.  
Although nova of faster speed class indeed expand faster, the average aspect ratio actually increases with speed class, which is opposite to the theoretical expectations. 

\begin{figure}
\centering
\includegraphics[width=1.0\linewidth]{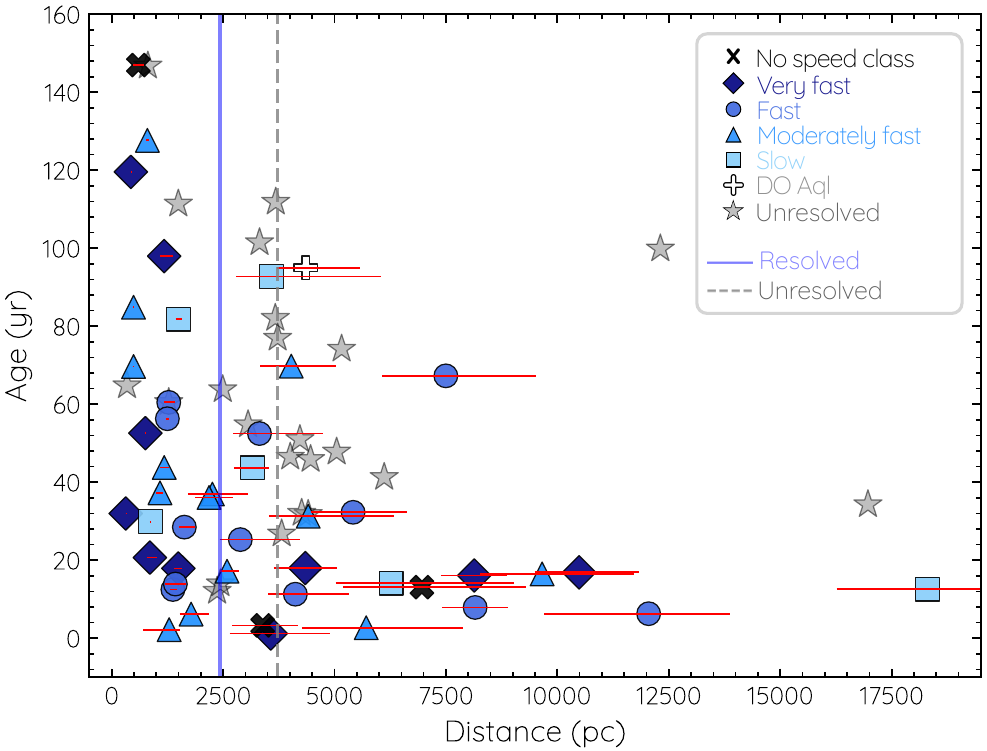}
\caption{
Distribution of age ($\tau$) with the distance ($d$) of the resolved and unresolved nova remnants. The 'Blues' points represent the resolved nova remnants in our sample, whereas gray stars are unresolved nova remnants. The median values of the distance for both samples (2420, 3720 pc for resolved and unresolved nova remnants, respectively) is shown by the vertical blue and gray lines.}
\label{fig:age}
\end{figure}

\section{Conclusions}\label{conclusions}

We present here the first optical imaging catalogue of nova remnants.  
The images have been obtained mostly through filters registering the H$\alpha$ or H$\alpha$+[N~{\sc ii}] $\lambda\lambda$6548,6584 \AA\ emission lines, with only one exception  obtained in the [O~{\sc iii}] $\lambda$5007 \AA\ emission line.

The catalogue shows a noticeable lack of available images in the archives of nova remnants along the direction of the Galaxy Center and Bulge where a significant fraction of the known Galactic novae can be found.  
We assume that the high extinction along these directions and stellar crowding has discouraged researchers from trying to image these nova remnants.

Out of the 66 nova remnants with images, 45 present extended emission and their nebular morphologies have been classified according to the ERBIAS scheme, which considers elliptical, round, bipolar, irregular, and asymmetric morphologies, as well as stellar (unresolved).  
Another five nova remnants have a G morphology, which are sources that, although unresolved, have a PSF that can be described by a 2D Gaussian wider than that of stars in the field of view.  
Most nova remnants are elliptical or round, with very few bipolar, particularly when compared to the known sample of PNe.  
This most likely indicates that the interaction time-scale of the nova ejecta with the accretion disk and binary companion is very short, much shorter than the wind-wind interaction in PNe.  

Compared to the general population of Galactic novae, resolved nova remnants are found preferentially at closer distances, as the lower extinction and larger H$\alpha$ surface brightness favour their detection.  
On the other hand, compared to unresolved nova remnants, resolved nova remnants have a higher detection rate at larger distances, in the range from 5 to 12 kpc, implying additional detection factors for the detection of extended emission such as the nova remnant age. 
Indeed, resolved nova remnants have a younger median age (33.2 yr) than unresolved nova remnants ($\approx$60 yr).  
About 80\% of nova remnants with age $\leq$60 yr are detected, but 50\% of older nova remnants miss detection.

The median size of the resolved nova remnants is 0.03 pc.  
The angular expansion of resolved nova remnants follows a relationship of 0.0725 pc per century, which implies a speed of 725 km~s$^{-1}$.  
Nova remnants of faster speed class expand at higher velocity, but the expected trend with decreasing aspect ratio is not found.  
The true aspect ratio of elliptical and round nova remnants follows a Gaussian distribution with dispersion 0.18.  
This implies that almost 70\% have a true aspect ratio smaller than 1.25, with less than 5\% having a true aspect ratio above 1.30.

It is unclear the difference between G and S nova remnants.  
They have similar mean distances and ages, but most of the former belong to a fast speed class, whereas about half of the sample of the latter are slow novae (V365\,Car, AR\,Cir, and CN\,Vel), or nova remnants older than 100 yr that might have already dispersed into the interstellar medium (AR\,Cir, V841\,Oph and CN\,Vel), or their images have a poor spatial resolution $\geq$0.03 pc above the median size of resolved nova remnants (V500\,Aql, RS\,Car, V365\,Car and RR\,Cha). 
Both observational biases and intrinsic properties may cause the lack of diffuse emission from S nova remnants.

\section*{Acknowledgements}

E.S. thanks UNAM DGAPA for funding as a posdoctoral researcher. E.S. and J.A.T. thank support from the UNAM PAPIIT project IN102324. 
G.R.-L. acknowledges support from CONAHCyT grant 263373 and Programa para el Desarrollo Profesional Docente (PRODEP) (Mexico). 
M.A.G.\ acknowledges financial support from grants CEX2021-001131-S funded by MCIN/AEI/10.13039/501100011033 and PID2022-142925NB-I00 from the Spanish Ministerio de Ciencia, Innovaci\'on y Universidades (MCIU) cofunded with FEDER funds. 
L.S acknowledges support from the UNAM PAPIIT project IN110122. 
This work as made extensive use of the NASA's Astrophysics Data System (ADS).

The Nordic Optical Telescope, operated on the island of La Palma jointly by Denmark, Finland, Iceland, Norway and Sweden, in the Spanish Observatorio del Roque de los Muchachos of the Instituto de Astrof\'{\i}sica de Canarias. We acknowledge the use of the ESO Science Archive Facility developed in partnership with the Space Telescope European Coordinating Facility (ST-ECF). We also acknowledge the ING archive, maintained as part of the CASU Astronomical Data Centre at the Institute of Astronomy, Cambridge, and finally, the STScI, operated by the Association of Universities for Research in
Astronomy, Inc., under NASA contract NAS5-26555.


\section*{DATA AVAILABILITY} 

The data used in this article can be found in public archives. The processed VLT MUSE data are available at the ESO Archive Science Portal\footnote{\url{https://archive.eso.org/scienceportal/home}} and the \textit{HST} data can be found in the Hubble Legacy Archive\footnote{\url{https://hla.stsci.edu}}. 
Otherwise the images can be provided by the author under reasonable request. 




\begin{thebibliography}{99}

\bibitem[Bailer-Jones et al.(2021)]{Bailer2021} 
Bailer-Jones, C.~A.~L., Rybizki, J., Fouesneau, M., et al.\ 2021, \aj, 161, 147. 

\bibitem[Banerjee et al.(2020)]{Banerjee+2020} 
Banerjee, D.~P.~K., Geballe, T.~R., Evans, A., et al.\ 2020, \apjl, 904, L23. 

\bibitem[Banerjee et al.(2023)]{Banerjee+2023} Banerjee, D.~P.~K., Evans, A., Woodward, C.~E., et al.\ 2023, \apjl, 952, L26. 

\bibitem[Bode(2010)]{Bode2010} Bode, M.~F.\ 2010, Astronomische Nachrichten, 331, 160. 

\bibitem[Bode \& Evans(2008)]{Bode2008} Bode, M.~F. \& Evans, A.\ 2008, Classical Novae, 2nd Edition. Edited by M.F. Bode and A. Evans. Cambridge Astrophysics Series, No. 43, Cambridge: Cambridge University Press, 2008.

\bibitem[Bond et al.(2024)]{Bond+2024} 
Bond, H.~E., Carter, C., Elmore, D.~F., et al.\ 2024, \aj, in press, arXiv:2409.06835. 

\bibitem[Bond \& Miszalski(2018)]{BM2018} 
Bond, H.~E. \& Miszalski, B.\ 2018, \pasp, 130, 094201. 

\bibitem[Castro Segura et al.(2021)]{CS+2021} 
Castro Segura, N., Knigge, C., Acosta-Pulido, J.~A., et al.\ 2021, \mnras, 501, 1951. 

\bibitem[Darnley et al.(2016)]{Darnley+2016} 
Darnley M.~J., Henze M., Bode M.~F., Hachisu I., Hernanz M., Hornoch K., Hounsell R., et al., 2016, \apj, 833, 149. 

\bibitem[\protect\citeauthoryear{Diaz et al.}{2018}]{Diaz+2018} 
Diaz M.~P., Abraham Z., Ribeiro V.~A.~R.~M., Beaklini P.~P.~B., Takeda L., 2018, \mnras, 480, L54. 

\bibitem[Downes \& Shara(1993)]{Downes+1993} Downes, R.~A. \& Shara, M.~M.\ 1993, \pasp, 105, 127. doi:10.1086/133139

\bibitem[Downes et al.(1997)]{Downes+1997} Downes, R., Webbink, R.~F., \& Shara, M.~M.\ 1997, \pasp, 109, 345. doi:10.1086/133900

\bibitem[Downes \& Duerbeck(2000)]{Downes+2000} Downes, R.~A. \& Duerbeck, H.~W.\ 2000, \aj, 120, 2007. doi:10.1086/301551

\bibitem[Downes et al.(2001)]{Downes+2001} Downes, R.~A., Webbink, R.~F., Shara, M.~M., et al.\ 2001, \pasp, 113, 764. doi:10.1086/320802

\bibitem[Duerbeck(1987)]{Duerbeck+1987} Duerbeck, H.~W.\ 1987, \ssr, 45, 1. doi:10.1007/BF00187826

\bibitem[Frew(2008)]{Frew2008} 
Frew, D.~J.\ 2008, Ph.D. Thesis

\bibitem[Gaposchkin(1957)]{Gaposchkin1957} 
Gaposchkin, C.~H.~P.\ 1957, Amsterdam, North-Holland Pub. Co.; New York, Interscience Publishers, 1957.

\bibitem[Gehrz et al.(1998)]{Gehrz1998} Gehrz, R.~D., Truran, J.~W., Williams, R.~E., et al.\ 1998, \pasp, 110, 3

\bibitem[Guerrero et al.(2025a)]{Guerrero+2025a}
Guerrero M.~A., Santamaria E., Liberato G., Parker Q.~A., Goncalves D.~R., Rodriguez-Gonzalez J.~B., Ritter A., et al., 2024, \aap, 694, A105. 

\bibitem[\protect\citeauthoryear{Guerrero et al.}{2025b}]{Guerrero+2025b} 
Guerrero M.~A., Santamaria E., Takeda L., Gonzalez-Carbajal J.~I., Cazzoli S., Ederoclite A., Toala J.~A., 2025, \aap, in press, arXiv, arXiv:2501.02501. 

\bibitem[\protect\citeauthoryear{Guerrero et al.}{2018}]{Guerrero+2018} 
Guerrero M.~A., Sabin L., Tovmassian G., Santamar{\'\i}a E., Michel R., Ramos-Larios G., Alarie A., et al., 2018, ApJ, 857, 80. 

\bibitem[Healy-Kalesh et al.(2024)]{HK+2024} 
Healy-Kalesh, M.~W., Darnley, M.~J., Harvey, {\'E}. J., et al.\ 2024, \mnras, 529, L175. 

\bibitem[Henden \& Munari(2007)]{Henden+2007} 
Henden, A. \& Munari, U.\ 2007, Information Bulletin on Variable Stars, 5803, 1

\bibitem[Hoffmann(2019)]{Hoffmann2019} 
Hoffmann, S.~M.\ 2019, \mnras, 490, 4194. 

\bibitem[I{\l}kiewicz et al.(2024)]{I2024+} 
I{\l}kiewicz, K., Miko{\l}ajewska, J., Shara, M.~M., et al.\ 2024, \apjl, 972, L14. doi:10.3847/2041-8213/ad6e5a

\bibitem[Kato \& Nakamura(2001)]{Kato+2001} 
Kato, T. \& Nakamura, Y.\ 2001, \iaucirc, 7559

\bibitem[Kholopov(1985)]{Kholopov+1985a}
Kholopov P.~N.\ 1985, okpz.book

\bibitem[Kholopov et al.(1985)]{Kholopov+1985b} 
Kholopov, P.~N., Samus, N.~N., Kazarovets, E.~V., et al.\ 1985, Information Bulletin on Variable Stars, 2681, 1

\bibitem[Kholopov et al.(1987)]{Kholopov+1987} 
Kholopov, P.~N., Samus', N.~N., Kazarovets, E.~V., et al.\ 1987, Information Bulletin on Variable Stars, 3058, 1

\bibitem[Lloyd et al.(1997)]{Lloyd1997} 
Lloyd, H.~M., O'Brien, T.~J., \& Bode, M.~F.\ 1997, \mnras, 284, 137

\bibitem[Lloyd et al.(1993)]{Lloyd1993} 
Lloyd, H.~M., Bode, M.~F., O'Brien, T.~J., et al.\ 1993, \mnras, 265, 457

\bibitem[Macfarlane et al.(2014)]{Macfarlane+2014} 
Macfarlane, S., Steeghs, D., \& Woudt, P.\ 2014, Stellar Novae: Past and Future Decades, 490, 115

\bibitem[\protect\citeauthoryear{Mikolajewska et al.}{1999}]{Mikolajewska+1999} 
Mikolajewska J., Brandi E., Hack W., Whitelock P.~A., Barba R., Garcia L., Marang F., 1999, \mnras, 305, 190. 

\bibitem[Mukai(2017)]{Mukai2017} 
Mukai, K.\ 2017, \pasp, 129, 062001

\bibitem[Munari et al.(2002)]{Munari+2002} 
Munari, U., Henden, A., Corradi, R.~M.~L., et al.\ 2002, Classical Nova Explosions, 637, 52. 

\bibitem[Mustel \& Boyarchuk(1970)]{Mustel1970} 
Mustel, E.~R. \& Boyarchuk, A.~A.\ 1970, \apss, 6, 183. 

\bibitem[O'Brien et al.(1994)]{OBrien1994} 
O'Brien, T.~J., Lloyd, H.~M., \& Bode, M.~F.\ 1994, \mnras, 271, 155

\bibitem[Orlando et al.(2017)]{Orlando2017} 
Orlando, S., Drake, J.~J., \& Miceli, M.\ 2017, \mnras, 464, 5003

\bibitem[\protect\citeauthoryear{Page et al.}{2022}]{Page+2022} 
Page K.~L., Beardmore A.~P., Osborne J.~P., Munari U., Ness J.-U., Evans P.~A., Bode M.~F., et al., 2022, MNRAS, 514, 1557. 

\bibitem[Pagnotta \& Schaefer(2014)]{Pagnotta2014} 
Pagnotta, A. \& Schaefer, B.~E.\ 2014, \apj, 788, 164. 

\bibitem[\protect\citeauthoryear{Parker et al.}{2006}]{Parker+2006} 
Parker Q.~A., Acker A., Frew D.~J., Hartley M., Peyaud A.~E.~J., Ochsenbein F., Phillipps S., et al., 2006, MNRAS, 373, 79. 

\bibitem[Ritter \& Kolb(2003)]{Ritter+2003} 
Ritter, H. \& Kolb, U.\ 2003, \aap, 404, 301. 

\bibitem[Schaefer(2022)]{Schaefer2022} 
Schaefer, B.~E.\ 2022, \mnras, 517, 6150. 

\bibitem[Sahman et al.(2015)]{Sahman+2015} 
Sahman, D.~I., Dhillon, V.~S., Knigge, C., et al.\ 2015, \mnras, 451, 2863. 

\bibitem[Sahman et al.(2018)]{Sahman+2018} 
Sahman, D.~I., Dhillon, V.~S., Littlefair, S.~P., et al.\ 2018, \mnras, 477, 4483. 

\bibitem[Saito et al.(2013)]{Saito+2013} 
Saito, R.~K., Minniti, D., Angeloni, R., et al.\ 2013, \aap, 554, A123. 

\bibitem[Santamar{\'\i}a et al.(2020)]{Santa+2020} 
Santamar{\'\i}a, E., Guerrero, M.~A., Ramos-Larios, G., et al.\ 2020, \apj, 892, 60. 

\bibitem[Santamar{\'\i}a et al.(2022)]{Santamaria+2022} 
Santamar{\'\i}a, E., Guerrero, M.~A., Toal{\'a}, J.~A., et al.\ 2022, \mnras, 517, 2567. 

\bibitem[Santamar{\'\i}a et al.(2024)]{Santamaria2024} Santamar{\'\i}a, E., Toal{\'a}, J.~A., Guerrero, M.~A., et al.\ 2024, \mnras, 530, 4531. 

\bibitem[Schaefer et al.(2022)]{Schaefer+2022} 
Schaefer B.~E., Walter F.~M., Hounsell R., Hillman Y., 2022, \mnras, 517, 3864. 

\bibitem[Schaefer(2022)]{Schaefer2022b} Schaefer, B.~E.\ 2022, \mnras, 517, 6150. 

\bibitem[Shafter(2002)]{Shafter2002} 
Shafter, A.~W.\ 2002, Classical Nova Explosions, 637, 462

\bibitem[Shara et al.(2017)]{Shara+2017} 
Shara, M.~M., I{\l}kiewicz, K., Miko{\l}ajewska, J., et al.\ 2017, \nat, 548, 558. 

\bibitem[Shara et al.(2024)]{Shara+2024a} 
Shara, M.~M., Lanzetta, K.~M., Garland, J.~T., et al.\ 2024, \mnras, 529, 212. 

\bibitem[Shara et al.(2024)]{Shara+2024b} 
Shara, M.~M., Lanzetta, K.~M., Garland, J.~T., et al.\ 2024, \mnras, 529, 224. 

\bibitem[Shara et al.(2024)]{2024arXiv} Shara, M.~M., Lanzetta, K.~M., Garland, J.~T., et al.\ 2024, arXiv:2410.12103. doi:10.48550/arXiv.2410.12103

\bibitem[Shara et al.(2007)]{Shara+2007} 
Shara, M.~M., Martin, C.~D., Seibert, M., et al.\ 2007, \nat, 446, 159. 

\bibitem[Shara et al.(2012)]{Shara+2012} 
Shara M.~M., Mizusawa T., Zurek D., Martin C.~D., Neill J.~D., Seibert M., 2012, \apj, 756, 107. 

\bibitem[Shara \& Moffat(1982)]{SM1982} 
Shara, M.~M. \& Moffat, A.~F.~J.\ 1982, \apjl, 258, L41. 

\bibitem[Starrfield et al.(2016)]{Starrfield2016} Starrfield, S., Iliadis, C., \& Hix, W.~R.\ 2016, \pasp, 128, 051001

\bibitem[Tappert et al.(2020)]{Tappert+2020} 
Tappert, C., Vogt, N., Ederoclite, A., et al.\ 2020, \aap, 641, A122. 

\bibitem[Tarasova(2015)]{Tarasova+2015} Tarasova, T.~N.\ 2015, Astronomy Reports, 59, 920. doi:10.1134/S1063772915090085

\bibitem[Truran \& Livio(1986)]{Truran1986} Truran, J.~W. \& Livio, M.\ 1986, \apj, 308, 721

\bibitem[Walder et al.(2008)]{Walder2008} Walder, R., Folini, D., \& Shore, S.~N.\ 2008, \aap, 484, L9

\bibitem[Wesson et al.(2008)]{Wesson+2008} Wesson, R., Barlow, M.~J., Corradi, R.~L.~M., et al.\ 2008, \apjl, 688, L21. 

\bibitem[Wolf et al.(2013)]{Wolf2013} Wolf, W.~M., Bildsten, L., Brooks, J., et al.\ 2013, \apj, 777, 136

\bibitem[Woudt et al.(2009)]{Woudt+2009} Woudt, P.~A., Steeghs, D., Karovska, M., et al.\ 2009, \apj, 706, 738. 

\end{thebibliography}




\clearpage

\appendix

\section{Comparison of Distance Scales}
\label{app.dist}

\begin{figure} 
\centering 
\includegraphics[width=1.0\columnwidth]{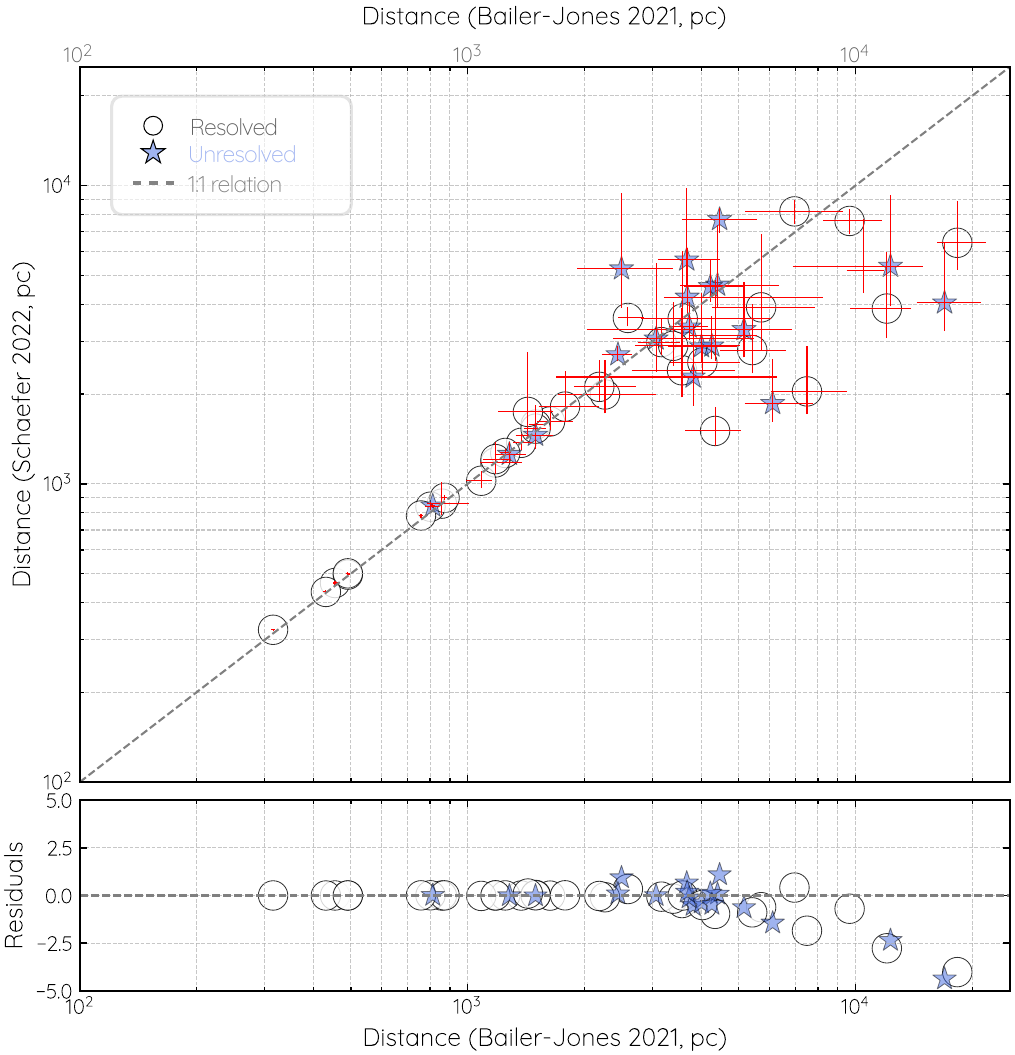}
\caption{Comparison of astrometric distances from Gaia DR3 distances \citep{Bailer2021} and other independent measurements reported by \citet{Schaefer2022b}. 
The gray dashed line shows the 1:1 relation. 
The white circles represent resolved nova remnants, whereas blue stars are unresolved nova remnants.
} 
\label{fig:dist} 
\end{figure}


The Gaia DR3 distances \citep{Bailer2021} towards the nova remnants with available images are compared to those based on independent measurements reported by \citet{Schaefer2022b} in Fig.~\ref{fig:dist}. 
In general, there is a strong correlation between Gaia distances and other independent measurements at distances closer than 3 kpc, with very slight deviations from the 1:1 line. 
At distances larger than 3 kpc, there is a considerable scatter around the 1:1 line, consistent with the expected decline in astrometric precision for more distant objects.  
Indeed most of these sources are within 3-$\sigma$ uncertainty of the 1:1 line (bottom panel of Fig.~\ref{fig:dist}).  
It can be noted that far resolved nova remnants (white circles) tend to have larger Gaia distances, a trend which is not completely followed by unresolved nova remnants (blue stars).  
It thus seems that at distances above 3 kpc the distances provided by \citet{Bailer2021} might overestimate the true distance.  
This highlights the importance of accounting for distance-dependent uncertainties in astrometric analyses and the relevance of distance measurements based on the expansion of resolved structures to provide critical comparisons. 

\section{
Related objects}
\label{app1}

\begin{table*}
\centering
\setlength{\columnwidth}{1.0\columnwidth}
\setlength{\tabcolsep}{1.6\tabcolsep}
\caption{Other novae remnants arranged by constellations and certainty classification.}
\label{tab:other_nova}
\resizebox{\textwidth}{!}{%
\begin{tabular}{llclccccc}
\hline
\hline
\multicolumn{1}{l}{Object} &
\multicolumn{1}{c}{$l, b$} & 
\multicolumn{1}{c}{$d_1$}  & 
\multicolumn{1}{c}{Image Epoch} & 
\multicolumn{1}{c}{Morphology} & 
\multicolumn{2}{c}{Radius}  &
\multicolumn{1}{c}{Age} & 
\multicolumn{1}{c}{Alternative nature} \\
\multicolumn{1}{c}{} &
\multicolumn{1}{c}{($^\circ$)} & 
\multicolumn{1}{c}{} & 
\multicolumn{1}{c}{} & 
\multicolumn{1}{c}{} & 
\multicolumn{1}{c}{(")}  &
\multicolumn{1}{c}{(pc)}  &
\multicolumn{1}{c}{(yr)} 
\\ 
\hline
\multicolumn{8}{c}{Ancient Nova Shells} \\
\hline
V1315\,Aql & 046.42$+$00.69 & $489\pm49$ & 2014 Aug & Em & 220 & 0.52 & 500 -- 1200 \\
Z\,Cam     & 141.38$+$32.63 & 214        & 2021 Nov & Em & 880, 2400 & 0.9, 2.5 & $2670^{+2100}_{-820}$ & $\dots$ \\
V1363\,Cyg & 070.92$+$00.91 & 1700       & IPHAS & E  &  50$\times$35 & 0.41$\times$0.29 & $\dots$ & $\dots$  \\
RX\,Pup    & 258.50$-$03.93 & 1600 & 2016 Dec & Em & 22, 120 & 0.17, 0.93 & 1300, 7000 & $\dots$ \\
\hline
\multicolumn{8}{c}{Nova Super Remnants} \\
\hline
KT\,Eri    & 207.99$-$32.02 & $5110^{+920}_{-430}$ & 2021 Oct -- 2022 Nov & E & 1080$\times$600 & 26.8$\times$14.9 & 25000 -- 51000 & $\dots$ \\
RS\,Oph    & 019.80$+$10.37 & $2400^{+300}_{-200}$ & IRAS & E & $2400\times720$ & $27.9\times8.4$ & $\dots$ & $\dots$ \\
\hline
\multicolumn{8}{c}{Dubious Nova Remnants} \\
\hline
V341\,Ara & 326.88$-$12.48 & 156.1$\pm$2.0 & 2004 May & I & $240\times180$ & $0.18\times0.14$ & 800 -- 1000 & Bow-shock nebula \\ 
BZ\,Cam   & 143.60$+$23.82 &   372$\pm$5   & 1996 Mar & I & $450\times75$ & $0.81\times0.14$ & $2700^{+2100}_{-800}$ & Bow-shock nebula \\ 
SY\,Can   & 210.01$+$36.44 & 405$\pm$5 & 2024 Apr-Jun & I & 450 & 0.88 & 7800 & Bow-shock nebula \\
Sco\,1437 & 343.30$-$00.66 & 2250      & 2012 Jul      & E         & $50\times46$ & $0.50\times0.46$ & 20000$\pm$2300 & PN \\
CK\,Vul   & 063.38$+$00.99 & 3200$^{+900}_{-600}$ & 2010 Jun & Bm & $36\times14$ & $0.56\times0.22$ & 355 & Luminous red nova \\
Te\,11    & 203.17$-$13.41 & $330\pm50$ & 2012 Feb & I-B & $44\times9$ & 0.070$\times$0.014 & 1540? & Bow-shock nebula \\
\hline
\end{tabular}
}
\vspace{0.15cm}
\end{table*}

The search for images of nova remnants has not included a number of nebulae otherwise typically associated with the nova remnant phenomenon.  
As listed in Tab.~\ref{tab:other_nova}, these are classified as the ancient nova shells V1315\,Aql, Z\,Cam, V1363\,Cyg, and RX\,Pup, and the nova super remnants KT\,Eri and RS\,Oph.  
It includes as well a number of dubious nova remnants, such as V341\,Ara, BZ\,Cam, SY\,Can, Sco\,1437, CK\,Vul, and Te\,11.  
The detailed properties of these sources are described below: \\
$\bullet$ \; V1315\,Aql is a 1 pc-sized shell around a nova-like variable of the SW\,Sex sub-type \citep{Sahman+2015}.  
Its ionized mass of $\simeq2\times10^{-4}$ M$_\odot$ is consistent with a nova shell, but its expansion velocity, $\simeq$25 km~s$^{-1}$, is atypically low \citep{Sahman+2018}.  
The nova age is estimated to be in the range 500-1200 years.  
It has been argued that the nova ejecta was originally much smaller, $\lesssim10^{-5}$ M$_\odot$, reducing its expansion velocity and increasing its mass as it swept up material from the surrounding medium. 
The ratio between the present ionized mass and original ejecta requires a medium density $\approx$0.01 cm$^{-3}$. 
\\
$\bullet$ \; Z\,Cam is an ancient shell nova around the dwarf nova of the same name \citep{Shara+2007}. 
Recent deep, large field of view images have revealed a set of multiple shells, with an outer shell with a physical radius up to 2.5 pc that would correspond to an ancient nova event occurring $\approx$2700 yr ago \citep{Shara+2024a}.  
The inner shell would have an age notably smaller, but still in the range around 1000 years \citep[more precisely 1300 yr according to][]{Shara+2012}. 
\\
$\bullet$ \; V1363\,Cyg has been proposed to be a 1 pc-sized shell around this dwarf nova, based on the inspection of shallow IPHAS H$\alpha$ and $r$ images \citep{Sahman+2015}. 
This detection is ambiguous and deserves further investigation. \\
$\bullet$ \; RX\,Pup is a symbiotic nova that experienced a classical nova outburst in the 1970s \citep{Mikolajewska+1999}.  
Deep H$\alpha$ and [O~{\sc iii}] imaging has revealed the presence of two arc-like features that can be interpreted as ancient nova shells \citep{I2024+}. \\
$\bullet$ \; KT\,Eri (Nova Eridani 2009) was recently proposed to be a recurrent nova with a recurrent period of a few decades \citep{Schaefer+2022}.  
This fact led to the search for large-scale emission around it that resulted in the discovery of a large, 50 pc-sized H$\alpha$-emitting shell \citep{Shara+2024b}.  
The enormous physical size, comparable to that of the nova super remnant M31-2008-12a in the Andromeda Galaxy \citep{Darnley+2016}, supports a nova super remnant nature for KT\,Eri. \\
$\bullet$ \; RS\,Oph is a symbiotic recurrent nova with a recurrence period of $\simeq$15 yr \citep{Page+2022}. 
The presence of a large nova super remnant shell has been proposed based on infrared \emph{IRAS} images \citep{HK+2024}. 
The detection of a shell and its association with RS\,Oph is dubious and deserves further investigation. \\
$\bullet$ \; CK\,Vul is a multi-bipolar nebula around a nova-like that has been historically classified as an old-slow nova \citep{SM1982}.  
It has been recently proposed to be a luminous red nova \citep{Banerjee+2020}, a very different outburst than that of a classical nova \citep{Munari+2002}.  \\
$\bullet$ \; HaTr\,5 is a round nebula proposed to be associated with the nova shell of Nova Sco\,1437 and the CV 2MASS\,J17012815$-$4306123  \citep{Shara+2017}.
A critical revision of ancient Chinese and Korean astronomy texts and proper motions of the CV \citep{Hoffmann2019} as well as the determination of a low expansion velocity of 27 km~s$^{-1}$ and an ionized mass too large for a nova remnant \citep{Guerrero+2025a} suggest it to be an unrelated planetary nebula (PN). \\
$\bullet$ \; Fr\,2-11, EGB\,4, and SY\,Can are bow-shock nebulae associated with the nova-like CV V341\,Ara \citep{Frew2008,BM2018,CS+2021}, the nova-like variable of VY Scl sub-type BZ\,Cam \citep{BM2018}, and the Z\,Cam-type CV SY\,Can \citep{Bond+2024}, respectively. 
All these nebulae have been proposed to be the result of a nova event, but these are most likely the result of the photoionization of H~{\sc ii} regions into the ISM or a wind-driven interaction with the ISM of high-speed CV systems.

\section{Comments on iconic nova remnants}
\label{app2}

$\bullet$V445\,Pup (Nova Puppis 2000) is the only known helium nova to date, identified on December 30, 2000, at a magnitude of $\sim$8.6 mag \citep{Kato+2001}. 
V445\,Pup is characterized by its absence of hydrogen and a gradual brightness decline resulting from substantial dust formation \citep{Banerjee+2023}. 
\cite{Woudt+2009} reported that a dust disk obscured a strongly bipolar shell with two rapidly moving knots as the tips of the bipolar lobes, as evidenced by post-outburst spectroscopy and imaging studies, particularly the multi-epoch integral field unit (IFU) spectroscopy presented by \cite{Macfarlane+2014}. 
The knots and the shell exhibit distinct spectral features, particularly the absence of He~{\sc i} recombination lines and strong forbidden emission lines, including [O~{\sc iii}] $\lambda\lambda$4959,5007 \AA\ and [O~{\sc ii}] $\lambda\lambda$7320,7330 \AA\ in the knots, which suggests shock heating or variations in elemental abundances.
The angular expansion of the knots is estimated to be 0.2 arcsec~yr$^{-1}$, in agreement with the 0.217 arcsec~yr$^{-1}$ previously reported by \cite{Woudt+2009}. 

$\bullet$V458\,Vul (Nova Vulpeculae 2007), discovered by Hiroshi Abe on August 8, 2007, is one of the few nova events known to have occurred inside a PN \citep{Wesson+2008}. 
The PN is described to present symmetrical bright features, termed "knots," around the central star. 
The nova progenitor has an apparent B magnitude of 18.2 mag \citep{Henden+2007}. 
\cite{Tarasova+2015} presented spectroscopic observations of the hybrid nova V458\,Vul, conducted from days 9 to 778 following the brightness peak, revealing short-period daily variations in the forbidden [Fe~{\sc vii}] line profiles during the nebular phase. 
The estimated abundances of helium, neon, and iron in the envelope are 4.4, 4.8, and 3.7 times greater than solar values, respectively. 
The mass of the envelope is determined to be 1.4$\times$10$^{-5}$~M$_{\odot}$.


\section{Stellar Remnants}

Narrow-band H$\alpha$ images of the sample of stellar nova remnants. 
In all plots, North is up, East is to the left. 
The first contour level in each frame is set at 5$\sigma$ above the background.


\begin{figure*} 
\centering 
\includegraphics[width=0.95\linewidth]{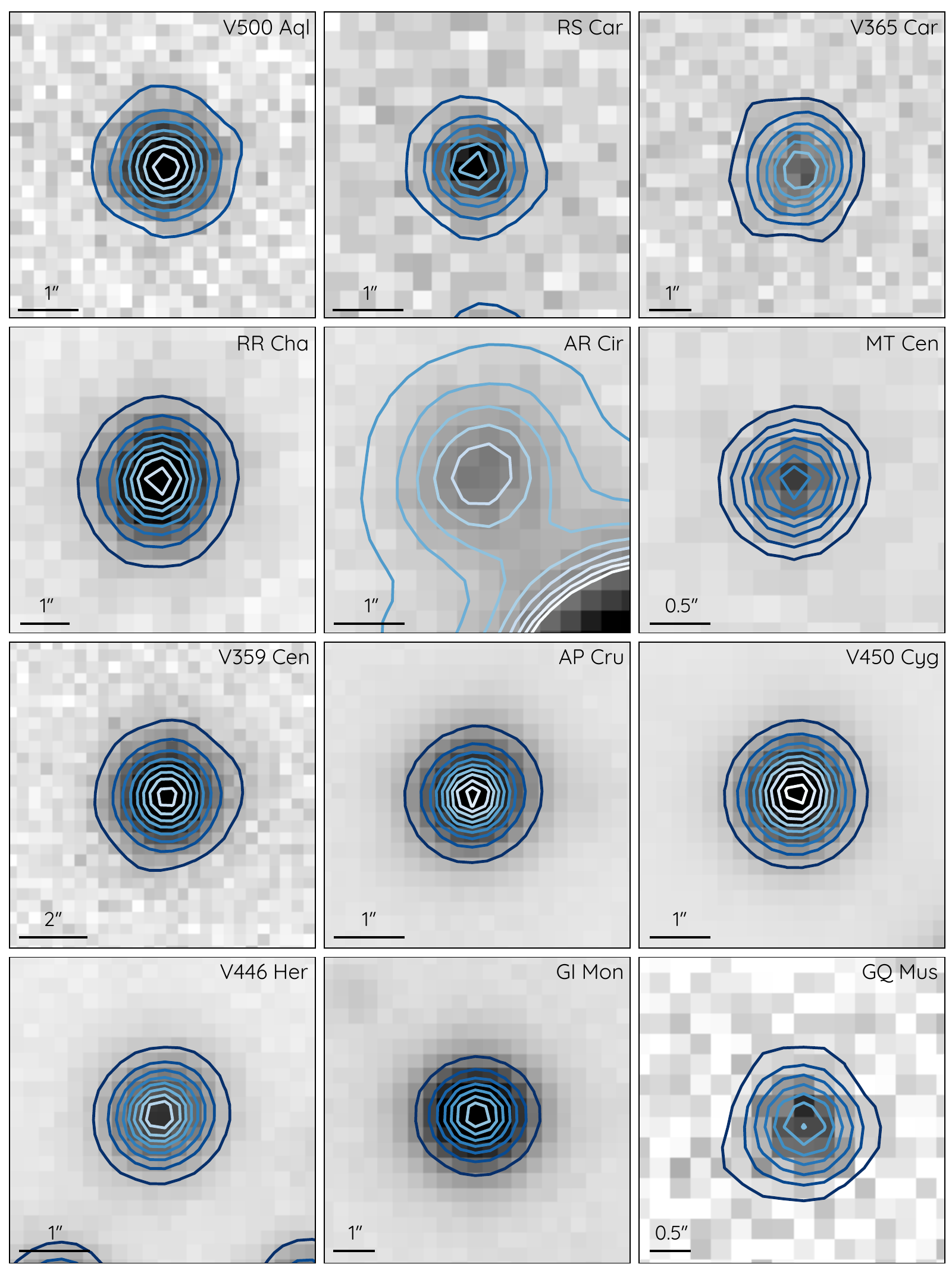}
\caption{Optical images of the nova stellar remnants V500\,Aql, RS\,Car, V365\,Car (top row), RR\,Cha, AR\,Cir, MT\,Cen (second raw), V359\,Cen, AP\,Cru, V450\,Cyg (third row), V446\,Her, GI\,Mon and GQ\,Mus (bottom row). North is up, East to the left.} 
\label{st1.img} 
\end{figure*}

\begin{figure*} 
\centering 
\includegraphics[width=0.95\linewidth]{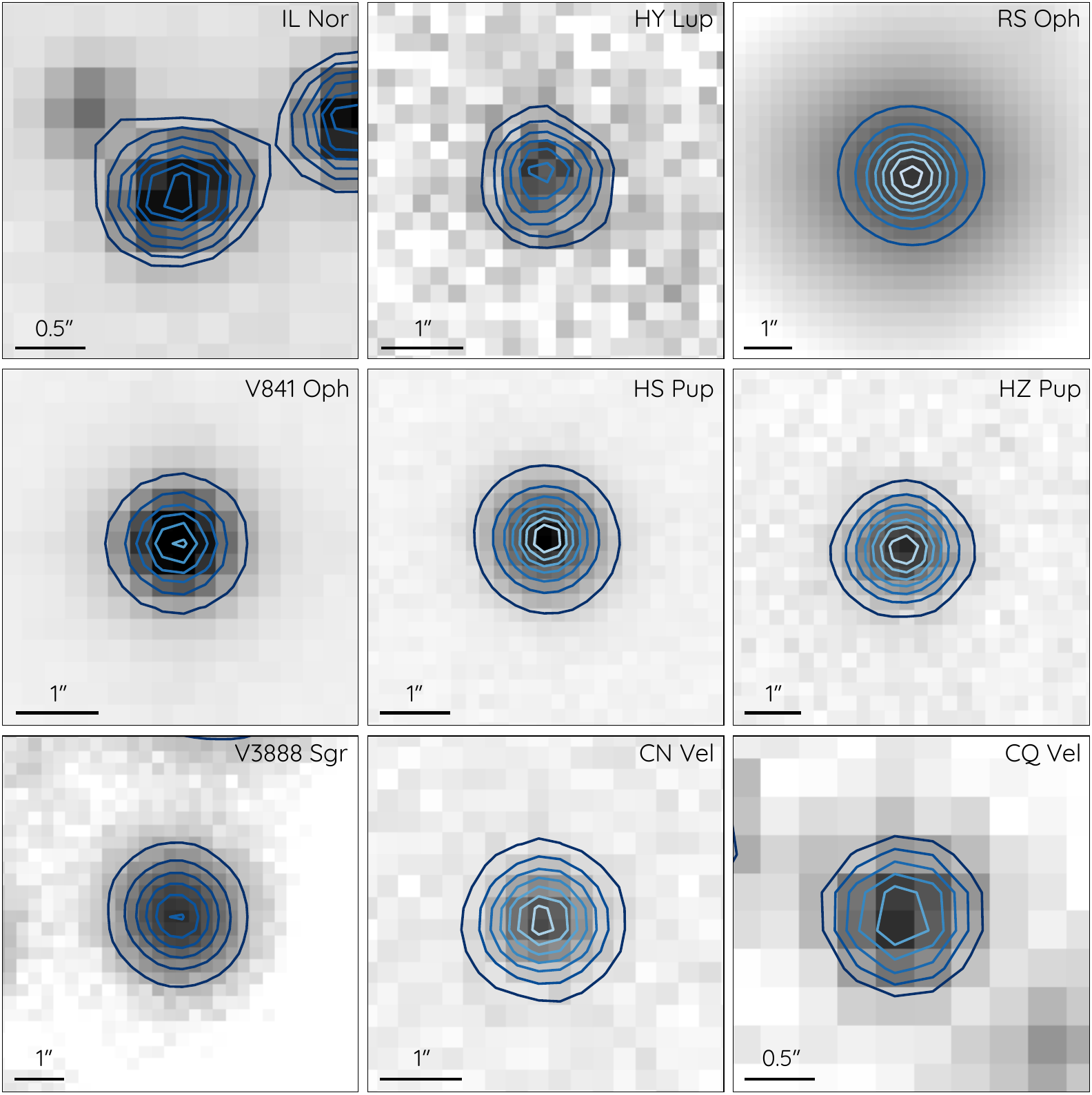}
\caption{Same as Fig.~\ref{st1.img} for the stellar remnants. Optical images of the nova stellar remnants, IL\,Nor, HY\,Lup, RS\,Oph (top row), V841\,Oph, HS\,Pup, HZ\,Pup (second raw), V3888\,Sgr, CN\,Vel and CQ\,Vel (bottom row). North is up, East to the left.} 
\label{st2.img} 
\end{figure*}


\bsp	
\label{lastpage}
\end{document}